# ULTRA-TUNING OF NONLINEAR DRUMHEAD MEMS RESONATORS BY ELECTRO-THERMOELASTIC BUCKLING

Ali Kanj[1], Paolo F. Ferrari[1], Arend M. van der Zande[1,2], Alexander F. Vakakis[1], Sameh Tawfick[1,3]

[1]Department of Mechanical Science and Engineering, University of Illinois at Urbana-Champaign, Illinois 61801, United States
[2]Materials Research Laboratory, University of Illinois at Urbana-Champaign, Illinois 61801, United States
[3]The Beckman Institute of Advanced Science and Technology, University of Illinois at Urbana-Champaign, Illinois 61801, United States

**Nonlinear micro-electro-mechanical systems (MEMS) resonators open new opportunities in sensing and signal manipulation compared to their linear counterparts by enabling frequency tuning and increased bandwidth. Here, we design, fabricate and study drumhead resonators exhibiting strongly nonlinear dynamics and develop a reduced order model (ROM) to capture their response accurately. The resonators undergo electrostatically-mediated thermoelastic buckling which tunes their natural frequency from 4.7 to 11.3 MHz, a factor of 2.4× tunability. Moreover, the imposed buckling switches the nonlinearity of the resonators between purely stiffening, purely softening, and even softening-to-stiffening. Accessing these exotic dynamics requires precise control of the temperature and the DC electrostatic forces near the resonator's critical-buckling point. To explain the observed tunability, we develop a one-dimensional physics-based ROM that predicts the linear and nonlinear response of the fundamental bending mode of these drumhead resonators. The ROM captures the dynamic effects of the internal stresses resulting from three sources: The residual stresses from the fabrication process, the mismatch in thermal expansion between the constituent layers, and lastly, the applied electrostatic forces. The ROM replicates the observed tunability of linear (within 5.5% error) and nonlinear responses even near the states of critical buckling. These remarkable nonlinear and large tunability of the natural frequency are valuable features for on-chip acoustic devices in broad applications such as signal manipulation, filtering, and MEMS waveguides.**

1. Introduction

Currently, micro-electro-mechanical systems (MEMS) resonators are indispensable for radio-frequency (RF) signal processing and filtering [1] [2], energy harvesting [3] [4] [5], and sensing and actuating [6] [7]. In the near future, MEMS resonators could enable megahertz to gigahertz acoustic signals that can couple to photonic [8] [9] and electric devices [10] [11]. Such coupling is promising for future applications like photonic-phononic memory [12] [13], quantum information control [14] [15], and integrated circuitry [16] [17]. For instance, acoustic waves offer an advantage over optical waves in inducing nonlinear interactions over a large space compared to the optical wavelength [18] [19], which enables nonreciprocal behaviors [20] [21] even in passive settings [22] [23] [24].

Buckled MEMS resonators result from the interplay between compressive stresses and the multi-physics arising at small scales, such as thermoelastic and electromechanical effects. While buckling presents a drawback in traditional structural design, a new paradigm exploits buckling to enable new applications in acoustic metamaterials [25]. For example, electrostatic and electrothermal buckling tunes the natural frequency [26] [27] [28] [29] and the geometric nonlinearity [30] [31] [32] [33] [34] in micro/nano-resonators. Moreover, buckling amplifies the weak thickness variabilities (<5%) among the unit cells of phononic waveguides until eventually breaking the stiffness-periodicity of the waveguides, which switches off the acoustic transmission [35].

In this work, we study the effect of buckling on MEMS drumhead resonators under different electrostatic and thermal conditions (section II). We experimentally report that electrostatic control of buckling tunes the natural frequency of micro-drumhead resonators by a factor >2.40× from 4.7 to 11.3 MHz (section III). We find that the maximal amount of electrostatic tunability depends on the internal stresses of the resonators set by the interplay of the device temperature and the fabrication-residual stresses. As for the nonlinear response, we switch the nonlinearity in the resonator's dynamics from softening to stiffening by electrothermally setting the resonators near the critical buckling, which is the state of minimum natural frequency (section IV). The realized stiffening near-critical buckling covers a broad band of frequencies, thus allowing for a wider operational range [36] [37]. The resonators switch from purely softening to purely stiffening by transitioning through a phase of softening-to-stiffening nonlinearity.

To explain the observed tunability, we develop a one-dimensional reduced-order model (ROM) that captures the essential mechanisms responsible for buckling in the resonators (section II). The proposed ROM lumps the resonator into a discrete mass. The mass translates along the direction of the bending motion while being subjected to the resonator's bending, stretching, thermal expansion, residual stresses, and the applied voltage on the resonator, which provokes buckling [27] [30]. These multi-physical effects are incorporated into the ROM by a discrete electrostatic force and two structural springs. These springs account for the resonator's bending and stretching by assuming an orthogonal configuration that mimics the von Mises truss model of buckling [38]. The free lengths of these springs are temperature-dependent to model the resonator's thermal and fabrication-residual stresses. Therefore, these temperature-dependent springs and the discrete electrostatic force induce compressive effects that buckle the ROM. Remarkably, although relatively simple, the derived discrete ROM predicts very accurately (within 5.5% error) the experimental tuning in natural frequency for the measured temperatures and electric voltages (section III). The ROM also captures the softening/stiffening switching in the nonlinearity of the response with an accuracy depending on the electrothermal conditions (section IV).

## 2. Experimental system and reduced-order model

Figure 1 describes the micro-drumhead resonators and reduced order model we used to study the effect of electrothermal buckling on the dynamics of resonators. Figures 1a-b show an optical image and schematic of the fabricated micro-drumhead resonators. Each resonator is formed of a ~60 nm thick circular silicon nitride ($SiN_x$) layer of ~10.1 µm diameter with a central hole of ~1.3 µm diameter over a vacuum gap of 150 nm. Patterned 60 nm thick gold electrodes on top of the $SiN_x$ layer allow electrostatic actuation. The underlying silicon is degenerately n++ doped to be conductive and serves as a global electrostatic gate. The $SiN_x$ is initially grown on top of a 150 nm Silicon Dioxide ($SiO_2$) sacrificial support layer. The $SiN_x$ layer is then released using hydrofluoric (HF) acid, where the central hole allows the HF to access and selectively etch the underlying sacrificial oxide. (cf. supplemental material section S1.a for additional information regarding the fabrication process).

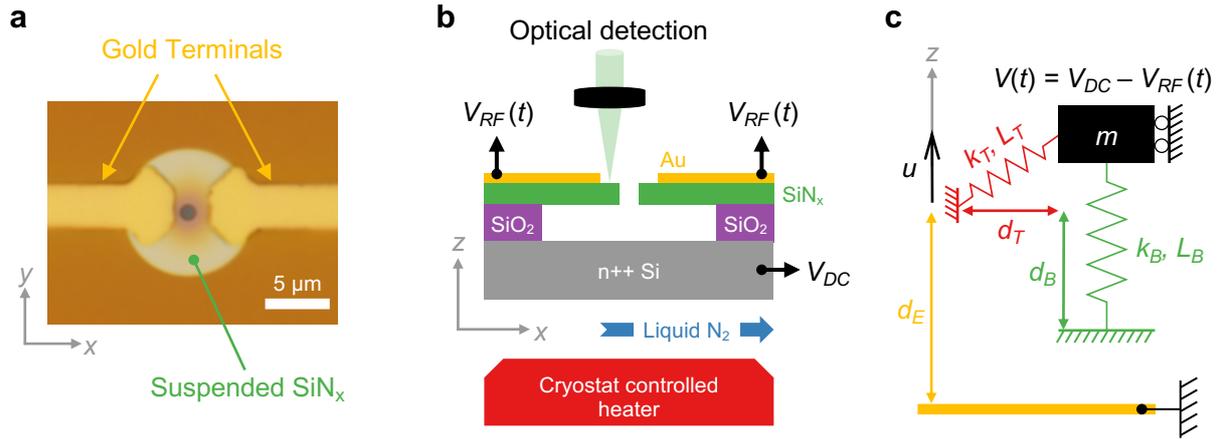

Fig. 1 Design and geometry of the micro-drumhead resonator. (a) Optical microscopic picture of the top view of the drumhead resonator at room temperature with $V_{dc} = 0$ V. The picture shows interferometric coloring (darker orange) at the center of the resonator due to the out-of-plane deflection resulting from buckling. (b) Schematic illustration of the resonator's cross-section passing through the centerline with the gold terminals. The schematic depicts the layered materials, the cryostat temperature control mechanism, the electrostatic actuation, and the optical detection apparatus. (c) Schematic of the reduced-order model (ROM) developed in this work to capture the electrothermal buckling effects on the dynamics of the resonator.

Figure 1b schematically shows the experimental measurement and tunable electrostatic and thermal controls. The samples are measured inside an optical cryostat, which allows electrical and optical inputs and thermal control. All measurements are performed in a vacuum, with pressure $< 2\times10^{-6}$ mBar. We electrostatically actuate the resonators by applying the radio frequency RF voltage $V_{RF}(t) = V_{amp} \sin(\Omega t)$ to the electrodes and a DC voltage $V_{DC}$ to the underlying silicon back gate, leading to a static out-of-plane force $F_{DC}$ (cf. (14)) which tunes the resonator's internal stresses and an RF force $F_{RF}$ (cf. (15)) which drives the resonator to resonate, discussed later as part of the reduced order model. To detect the motion, we focus a laser near the center of the resonator and measure the time-varying reflected light using a photodiode attached to a network analyzer. The resonator is partially transparent, so the light reflecting from the back-gate and the resonator's surface interferes due to Fabry-Perot

Interferometry. As a result, the change in reflected light intensity is proportional to the amplitude of motion.

To unravel the relative contributions from electrostatic tuning and the resonator's internal stress, we independently control the internal stress by adjusting the temperature in the optical cryostat. This temperature-change tunes the internal stress in the resonator by the thermal expansions/contractions induced within the structural constraints. As shown in Supplementary Figure S1, when under compression, induced thermal stresses cause the resonator to buckle and deflect out of the plane, creating interferometry fringes depending on their buckled profile. We should note that the interplay between these thermoelastic effects and the fabrication-induced residual stresses define the internal stress (both tensile and compressive) for each temperature. Since the residual stresses are fixed in our experiments, we control the temperature to vary the internal stress.

Before discussing experimental results, we illustrate each of these multi-physics phenomena (i.e., electrostatic force, thermal stresses, and fabrication-induced residual stresses) using the reduced-order model (ROM). Fig. 1c shows the ROM along with the relations of all its constitutive elements. The proposed ROM lumps the drumhead resonator into a discrete mass $m$. As shown by the rollers in Fig. 1c, the mass $m$ translates with a displacement coordinate $u$ in the z-direction along which the flexural vibrations of the resonator occur while being subjected to the resonator's bending, stretching, thermal expansion, residual stresses, and the applied voltage on the resonator, which provokes buckling [39] [30] [27] [35]. Our ROM captures all these multi-physics effects in the form of discrete springs and forces connected and applied to the translating mass. The configuration of the springs in the ROM is inspired by the von Mises truss model of bistable buckling phenomena [38] by attaching a stretching spring orthogonally to the bending direction.

In addition, the ROM considers the thermal and fabrication-residual stresses by imposing temperature-dependent compressions in the assumed springs. Specifically, the ROM lumps the effective inertia of the resonator into a rigid mass. Therefore, the mass $m$ represents the effective mass of the resonator during its fundamental mode of vibrations, which constitutes the mode of interest for this ROM. The ROM captures the bending and stretching of the resonator by the springs of stiffnesses, $k_B$ and $k_T$, respectively. The bending spring $k_B$ aligns with the z-direction whereas the stretching spring $k_T$ elongates diagonally with the translation of $m$. For example, in the absence of all forces and springs in the ROM other than the stretching spring, the zero-equilibrium position (i.e., $u = 0$ m) will be unstable if the stretching spring $k_T$ is compressed to fit within the distance $d_T$. In this case, $m$ snaps from this unstable zero-equilibrium position to another stable equilibrium position whose location depends on the amount of precompression in the stretching spring $k_T$. Hence, it is important to quantify the value of precompression in the stretching spring, which we do in the ROM by defining the parameter $\delta_T$ as,

$$\delta_T \stackrel{\text{def}}{=} \frac{L_T - d_T}{d_T} \tag{1}$$

where $L_T$ is the free-length of the stretching spring that is confined within the distance $d_T$. With the definition in (1), the force exerted by the stretching spring in the z-direction is found to be:

$$F_T = -k_T u \left[ 1 - \frac{1 + \delta_T}{\sqrt{1 + \left(\frac{u}{d_T}\right)^2}} \right]. \tag{2}$$

Now, under the single effect of the stretching force $F_T$ in (2), the mass $m$ attains the stable equilibrium $u_{EQM}$ approximated to $\mathcal{O}\left[\left(\frac{u_{EQM}}{d_T}\right)^{\frac{3}{2}}\right]$ by:

$$\begin{cases} u_{EQM} = 0 \text{ for } \delta_T \leq 0 \text{ (i.e., spring } k_T \text{ under pretension)} \\ u_{EQM} = \pm d_T \sqrt{\frac{2\delta_T}{1+\delta_T}} \text{ for } \delta_T > 0 \text{ (i.e., spring } k_T \text{ under precompression)} \end{cases}. \tag{3}$$

Equation (3) shows that $\delta_T$ determines the stable equilibrium of mass $m$, which represents the in-plane strains in the resonator as defined in (1). These in-plane strains result during the fabrication process and by thermoelastic deformation of the resonator; therefore, $\delta_T$ depends on the fabrication-residual strains and the resonator's temperature. However, the fabrication and thermal strains do not only affect the in-plane strains in the resonator but also lead to mismatches in deformations/strains between the different material layers that anchor and form the resonator (cf. Fig. 1b). These strain mismatches, especially between the layers of material around the anchoring boundary of the resonator, induce moments that bend the resonator in the transverse z-direction of Fig. 1.

To account for the bending effect of the strain mismatches, we characterize the bending spring $k_B$ in the ROM with the parameter,

$$\delta_B \stackrel{\text{def}}{=} \frac{L_B - d_B}{d_B} \tag{4}$$

where $L_B$ corresponds to its free-length and $d_B$ corresponds to the distance confining the bending spring when $u = 0$ m as shown in Fig. 1c. The definition of $\delta_B$ in (4) leads to a bending force $F_B$ expressed as:

$$F_B = -k_B d_B \left(\frac{u}{d_B} - \delta_B\right) \tag{5}$$

Like $\delta_T$, $\delta_B$ depends on the fabrication-residual strains and the temperature $T$ of the resonator. We determine these temperature dependences of $\delta_T(T)$ and using experimental measurements that we run at different temperatures controlled by the cryostat depicted in Fig. 1b. Based on these experiments (cf. supplemental material section S3.a and Fig. S14a-b) and literature models [40], we find that $\delta_T$ and $\delta_B$ scale with temperature $T$ as,

$$\delta_T(T) = \theta_0 + \theta_1 T \tag{6a}$$

$$\delta_B(T) = \beta_0 + \beta_1 T + \beta_2 T^2, \tag{6b}$$

where the coefficients $\theta_0$, $\theta_1$, $\beta_0$, $\beta_1$, and $\beta_2$ are independent of temperature and voltage. These coefficients are dictated by the structural properties of the resonator (e.g., geometry and material) and the fabrication-residual stresses.

To model the electrostatic effects, the ROM replaces the distributed electrostatic force applied to the gold terminals (Fig. 1a-b) with an effective electrostatic point force $F_E$ exerted at the mass $m$ of Fig. 1c [41]. To compute $F_E$, we assume that $m$ in the ROM experiences a voltage drop $V(t) = V_{DC} - V_{RF}(t)$ between the gold terminals and the Si ground (cf. Fig. 1b). We apply a simplification assuming that $m$ and the electric ground in Fig. 1c form a parallel plates capacitor with a gap of $d_E + u$ and overlaying surfaces of $S$, which lead to the estimate,

$$F_E = -\frac{\epsilon_0 S}{2} \cdot \frac{V^2}{(d_E + u)^2} \tag{7}$$

where $\epsilon_0$ is the permittivity of free space. The expression in (7) lumps into $F_E$ the total effect of the distributed electrostatic load exerted on the resonator in Fig. 1a.

Since the resonator undergoes bending, the gap between the Si ground and the deformed resonator is not uniformly equal to $d_E + u$ as dictated in (7). To account for the gap variation along the resonator, we consider the deformation of the resonator according to the fundamental mode, which leads to a pre-factor $c(u) \stackrel{\text{def}}{=} \frac{\epsilon_0 S}{2}$ such that,

$$F_E = -\frac{c(u)}{(d_E + u)^2} V^2 \approx -\frac{c_0 + c_1 u}{(d_E + u)^2} V^2 \tag{8}$$

where $c_0$ and $c_1$ are constant coefficients for all voltages and temperatures (cf. supplemental material section S4.a and Fig. S15 for the derivation of (8)). With the forces in (2), (5), and (8), the mass in the ROM of Fig. 1c is subjected to the following external force along the z-direction:

$$F_{ext}(u; T, V) = F_B + F_T + F_E \tag{9}$$

$$= -\left\{k_B d_B\left(\frac{u}{d_B}-\delta_B\right)+k_T u\left[1-\frac{1+\delta_T}{\sqrt{1+\left(\frac{u}{d_T}\right)^2}}\right]+\frac{c_0+c_1 u}{(d_E+u)^2}V^2\right\}.$$

To generalize the ROM for different device geometries, we nondimensionalize the displacements and the forces by selecting the distance $d_B$ and the stiffness $k_B$ as references for normalization, yielding the following nondimensional external force,

$$\bar{F}_{ext}(\bar{u};T,V) \stackrel{\text{def}}{=} \frac{F_{ext}}{k_B d_B}$$

$$= -\left\{\bar{u}-\delta_B(T)+\kappa_T \bar{u}\left[1-\bar{d}_T\frac{1+\delta_T(T)}{\sqrt{\bar{d}_T{}^2+\bar{u}^2}}\right]+\frac{\gamma_0+\gamma_1 \bar{u}}{(\bar{d}_E+\bar{u})^2}V^2\right\} \quad (10)$$

where $\bar{u} \stackrel{\text{def}}{=} \frac{u}{d_B}, \bar{d}_T \stackrel{\text{def}}{=} \frac{d_T}{d_B}, \kappa_T \stackrel{\text{def}}{=} \frac{k_T}{k_B}, \bar{d}_E \stackrel{\text{def}}{=} \frac{d_E}{d_B}$, and $\gamma_0 + \gamma_1 \bar{u} = \frac{c_0}{2k_B d_B{}^3}+\frac{c_1}{2k_B d_B{}^3}u = \frac{\epsilon_0 S}{2k_B d_B{}^3}$. Note that the nondimensionalization in (10) reduces the number of involved parameters in the ROM by two, namely, the reference distance $d_B$ and stiffness $k_B$.

Recall that in the experiments, we impose the temperature $T$ of the resonator and the DC voltage $V_{DC}$, and then we study the effect of these conditions on the vibration of the resonator as depicted in Fig. 2. Alternatively in the ROM, we can identify the effect of $T$ and $V_{DC}$ using (10) by computing the equilibrium normalized displacement $\bar{u}_{EQM}(T,V_{DC})$ that sets $\bar{F}_{ext}$ to zero when $V = V_{DC}$ (i.e., imposing Newton's first law). Since the condition $\bar{F}_{ext} = 0$ can admit many solutions, we select $\bar{u}_{EQM}$ as the zero of $\bar{F}_{ext}$ with the strongest stability, which is the zero value with the largest value of $\left(-\frac{d\bar{F}_{ext}}{d\bar{u}}\right) > 0$. We refer to $\bar{u}_{EQM}(T,V_{DC})$ as the DC equilibrium which can be mathematically expressed as:

$$\bar{F}_{ext}(\bar{u}_{EQM};T,V_{DC}) = 0 \text{ with maximum } \left(-\frac{d\bar{F}_{ext}}{d\bar{u}}\right)\Big|_{\bar{u}=\bar{u}_{EQM}} > 0. \quad (11)$$

To study the vibrations using the ROM, we perturb the DC equilibrium by applying a weak RF voltage such that $V_{RF}{}^2 \ll 2V_{RF}V_{DC}$. This assumption is consistent with our measurements where we apply a maximum RF voltage of 0.316 $V_p$ (peak value) with a smallest DC voltage of 5 V. The assumption for weak AC voltage allows us to neglect the $V_{RF}^2$ term in $\bar{F}_{ext}$ resulting from $V = V_{DC} + V_{RF}(t)$ in (10), which leads to the form,

$$\bar{F}_{ext}\big(\bar{u}; T, V_{DC} + V_{RF}(t)\big)$$

$$\approx -\left\{\bar{u} - \delta_B(T) + \kappa_T \bar{u}\left[1 - \bar{d}_T \frac{1 + \delta_T(T)}{\sqrt{\bar{d}_T^2 + \bar{u}^2}}\right] + \frac{\gamma(\bar{u})}{(\bar{d}_E + \bar{u})^2}\big[V_{DC}^2 + 2V_{DC}V_{RF}(t)\big]\right\} \quad (12a)$$

$$\Rightarrow \bar{F}_{ext}\big(\bar{u}; T, V_{DC} + V_{RF}(t)\big) \approx \bar{F}_{ext}(\bar{u}; T, V_{DC}) - 2\frac{\gamma(\bar{u})}{(\bar{d}_E + \bar{u})^2} V_{DC}V_{RF}(t) \quad (12b)$$

where $\gamma(\bar{u}) \stackrel{\text{def}}{=} \gamma_0 + \gamma_1 \bar{u}$.

Defining a perturbative (dimensional) displacement $w\big(t; T, V_{DC} + V_{RF}(t)\big) \stackrel{\text{def}}{=} u\big(t; T, V_{DC} + V_{RF}(t)\big) - u_{EQM}(T, V_{DC})$, we write the equation of motion by applying Newton's second law to the ROM such that,

$$m \frac{d^2 w}{dt^2} = k_B d_B \bar{F}_{ext}\big(\bar{u}_{EQM} + \bar{w}; T, V_{DC} + V_{RF}(t)\big) \quad (13a)$$

$$\xRightarrow{\text{applying (12b)}} \frac{m}{k_B} \cdot \frac{d^2 \bar{w}}{dt^2} - \bar{F}_{ext}\big(\bar{u}_{EQM} + \bar{w}; T, V_{DC}\big) = -2 \frac{\gamma(\bar{u})}{(\bar{d}_E + \bar{u})^2} V_{RF} V_{DC}(t) \quad (13b)$$

where $\bar{w} \stackrel{\text{def}}{=} \frac{w}{d_B}$ and $t$ is the time variable. Note that (13a-b) do not incorporate damping because we focus on the detuning and the nonlinearity of the frequency response of the resonator. Normalizing the time variable by $\tau \stackrel{\text{def}}{=} \omega_B t$ where $\omega_B \stackrel{\text{def}}{=} \sqrt{\frac{k_B}{m}}$ and defining $\bar{F}_{DC}(\bar{u}) \stackrel{\text{def}}{=} -\bar{F}_{ext}(\bar{u}; T, V_{DC})$, we write (12b) as,

$$\frac{d^2 \bar{w}}{d\tau^2} + \bar{F}_{DC}\big(\bar{u}_{EQM} + \bar{w}\big) = -2 \frac{\gamma(\bar{u}_{EQM} + \bar{w})}{(\bar{d}_E + \bar{u}_{EQM} + \bar{w})^2} V_{DC} V_{amp} \sin\left(\frac{\Omega}{\omega_B} \tau\right), \quad (14)$$

which forms the equation of motion of the undamped resonator forced by the sinusoidal voltage $V_{RF}(t) = V_{amp} \sin(\Omega t)$ with $V_{amp}^2 \ll 2 V_{amp} V_{DC}$. Approximating the dynamics up to the order $\mathcal{O}(\overline{w}^5)$, and defining $\Gamma(\overline{u}) \stackrel{\text{def}}{=} \frac{\gamma(\overline{u}_{EQM}+\overline{w})}{(\overline{d}_E+\overline{u}_{EQM}+\overline{w})^2}$, $\kappa_n \stackrel{\text{def}}{=} \frac{1}{n!}\frac{d^n \overline{F}_{DC}}{d\overline{u}^n}\Big|_{\overline{u}=\overline{u}_{EQM}}$, and $\Gamma_n \stackrel{\text{def}}{=} \frac{2}{n!}\frac{d^n \Gamma}{d\overline{u}^n}\Big|_{\overline{u}=\overline{u}_{EQM}}$,

we obtain:

$$\frac{d^2\overline{w}}{d\tau^2} + \sum_{n=1}^{5}\left[\kappa_n + \Gamma_n \sin\left(\frac{\Omega}{\omega_B}\tau\right)\right]\overline{w}^n = \overbrace{-2\Gamma(\overline{u}_{EQM})V_{DC}V_{amp}\sin\left(\frac{\Omega}{\omega_B}\tau\right)}^{\overline{F}_{RF}(t)}. \quad (15)$$

We note from (15) that the applied RF voltage induces not only an external excitation $\overline{F}_{RF}(t)$ but also parametric excitations due to the $\Gamma_n \sin\left(\frac{\Omega}{\omega_B}\tau\right)$ terms (on the left-hand side). However, after identifying the parameters of the ROM with the experimental system as explained in the next section, we compare the $\Gamma_n \sin\left(\frac{\Omega}{\omega_B}\tau\right)$ term to the respective constant ($\kappa_n$) term on the left-hand side of (15), and we find that the parametric excitation terms are negligible (as explained in the supplemental material section S4.b). Hence, we ignore the parametric excitation in the system and assume the following equation,

$$\frac{d^2\overline{w}}{d\tau^2} + \kappa_1\overline{w} + \kappa_2\overline{w}^2 + \kappa_3\overline{w}^3 + \kappa_4\overline{w}^4 + \kappa_5\overline{w}^5 = -2\Gamma(\overline{u}_{EQM})V_{DC}V_{amp}\sin\left(\frac{\Omega}{\omega_B}\tau\right), \quad (16)$$

which approximates the dynamics up to the order $\mathcal{O}(\overline{w}^5)$. Nevertheless, equation (16) accounts only for the electrostatic and stretching nonlinearity of the ROM in Fig. 1c (i.e., the $\kappa_n$ terms) and disregards the geometric nonlinearity due to bending. Therefore, we add a bending nonlinear term $\kappa_3^B$ to (16) leading to:

$$\frac{d^2\overline{w}}{d\tau^2} + \kappa_1\overline{w} + \kappa_2\overline{w}^2 + (\kappa_3 + \kappa_3^B)\overline{w}^3 + \kappa_4\overline{w}^4 + \kappa_5\overline{w}^5 = -2\Gamma(\overline{u}_{EQM})V_{DC}V_{amp}. \quad (17)$$

We decide to approximate up to the quintic order of $\overline{w}$ (i.e., $\mathcal{O}(\overline{w}^5)$) because the nonlinear response in our measurements exhibits a stiffening-to-softening behavior as shown in Figs. 5b and 4e [30]. Such stiffening-to-softening behavior necessitates nonlinearities of at least quartic order (i.e., $\mathcal{O}(\overline{w}^4)$) [30]. To this end, the frequency detuning along the backbone curve of the forced response is expressed as,

$$\frac{\sigma}{\omega_B} = \frac{3}{8\sqrt{\kappa_1}} \left( \kappa_3^B + \underbrace{\kappa_3 - \frac{10}{9}\frac{\kappa_2^2}{\kappa_1}}_{\kappa_3^{T+E}} \right) \bar{w}_{amp}^2 + \left( -\frac{7}{8}\frac{\kappa_2 \kappa_4}{\sqrt{\kappa_1}^3} + \frac{5}{16}\frac{\kappa_5}{\sqrt{\kappa_1}} \right) \bar{w}_{amp}^4, \qquad (18)$$

where we denote by $\sigma \stackrel{\text{def}}{=} \omega_{Peak} - \sqrt{\kappa_1}$ the shift in peak-frequency between the nonlinear and the linear systems, by $\bar{w}_{amp}$ the amplitude of steady-state oscillations, and by $\kappa_3^{T+E}$ the effective cubic nonlinearity resulting from the stretching and the electrostatic effects in the ROM of Fig. 1c. Thus, for a stiffening-to-softening behavior, the coefficient of $\bar{w}_{amp}^2$ in (18) must be positive to stiffen the response at moderate amplitudes, and the coefficient of $\bar{w}_{amp}^4$ must be negative to soften the response at strong amplitudes.

## 3. Electrostatic frequency tuning mediated by thermoelastic buckling

In Figure 2, we explore the relative contributions of temperature and electrostatic forces on the tuning of the linear response of the drumhead resonators. Figs. 2a, 2b, and 2c show the linear frequency response for low drive amplitudes $V_{RF}$ gathered from the same resonator for increasing values of $V_{DC}$ at 340 K, 320 K, and 295 K, respectively. In Fig. 2, the recorded $V_{Meas}$ corresponds to the peak value of the amplitude of RF voltage measured by the photodetector depicted in Fig. 1b. The $V_{Meas}$ is proportional to the steady-state amplitude of oscillation $w_{amp}$ of the resonator via a constant labeled $\alpha$ (i.e., $w_{amp} = \alpha V_{Meas}$) based the employed Fabry-Pérot interferometry detection scheme (cf. supplemental material section S1.b). We note that $\alpha$ sensitively depends on the interferometric gap size and will be different for different values of temperature and $V_{DC}$.

The measurements in Figure 2 reveal that the temperatures and DC voltage drastically affect the natural frequency of the resonator. For instance, at all temperatures of Fig. 2, the natural frequency presents two regimes of tunability with increasing $V_{DC}$. In the first regime of increasing $V_{DC}$, subfigures (i) at each temperature, the natural frequency of the resonator

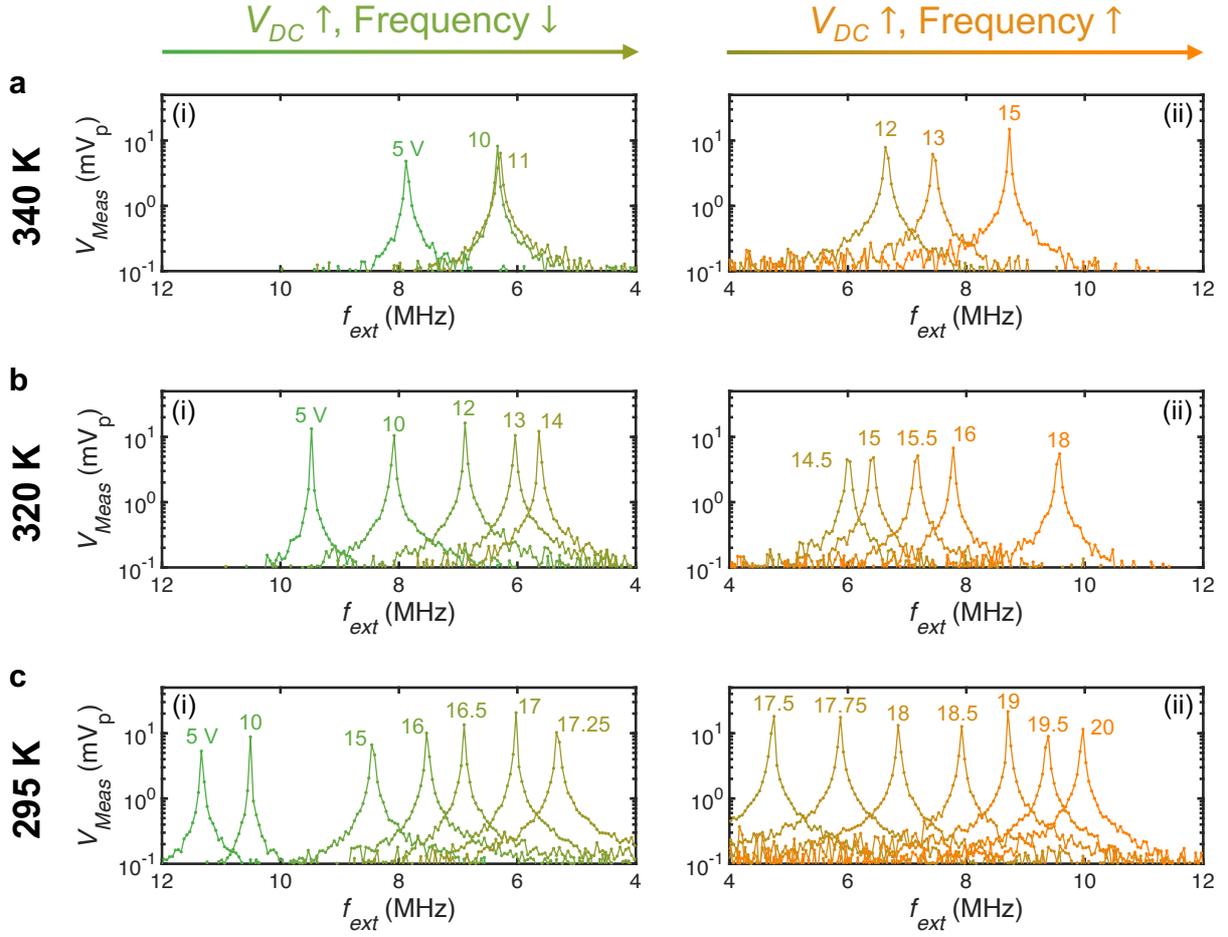

Fig. 2. Electrostatic and thermoelastic tunability of the resonator's linear dynamic response. Experimentally measured frequency response of the resonator at a temperature of (a) 340 K, (b) 320 K, and (c) 295 K for increasing $V_{DC}$ whose values label the respective respoonse. For each temperature in Fig. 2a-c, the increase in $V_{DC}$ provokes two tunability-regimes where the natural frequency either decreases or increases as shown in (i) and (ii), respectively. The voltage $V_{Meas}$ measured by the optical detection apparatus is proportional to the steady-state displacement of the resonator. The plots only show the response in a forward (increasing) sweep of the frequency $\Omega = 2\pi f_{ext}$ of the applied $V_{RF}(t) = V_{amp}\cos(2\pi f_{ext} t)$. The voltage unit $V_p$ denotes that the RF voltages are reported using their peak (amplitude) values. All the displayed responses are recorded for $V_{amp} = 3.16$ mV$_p$, except for $V_{DC} = 5$ V in Fig. 2a,c where $V_{amp} = 10$ mV$_p$.

*decreases*, whereas in the second regime of increasing $V_{DC}$, in the subfigures (ii), the natural frequency *increases*.

To further highlight and observe the tunability, Fig. 3a plots the natural frequency of the resonator versus $V_{DC}$ at 370 K, 360 K, 340 K, 320 K, and 295 K. The experimentally measured values are represented as points, while the lines indicate the trends predicted by the proposed ROM. The natural frequency at all these temperatures exhibits the two mentioned regimes of tunability with $V_{DC}$, which manifest in an experimental local minimum for each temperature value except for 370 K. However, this tunability with $V_{DC}$ becomes stronger at smaller temperatures for the measured resonator, as demonstrated by the sharp minimum frequency cusp. We observe that the natural frequency exhibits an increase from 7.2 MHz for $V_{DC} = 2.5$ V to 9.9 MHz for $V_{DC} = 15$ V at 370 K (~1.3× increase) versus an increase from 4.7 MHz for $V_{DC} = 17.5$ V to 11.3 MHz for $V_{DC} = 5$ V at 295 K (~2.4× increase). This large tunability observed at smaller temperatures (e.g., at 295 K and 320 K in Fig. 3a) is repeatable and observed in additional resonators fabricated as shown in the supplemental material section S2 and Fig. S2-S13.

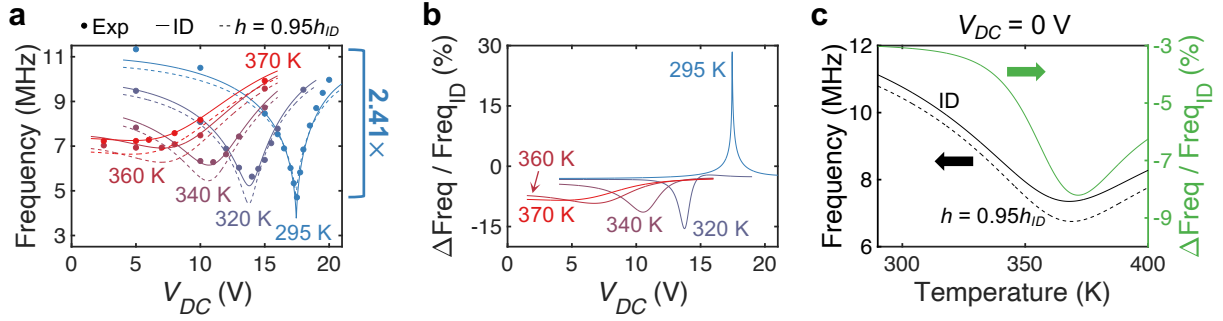

Fig. 3. Large electro-thermoelastic tunability of the resonator's natural frequency. (a) The natural frequency of the resonator as a function of the applied $V_{dc}$ for five different temperatures. The round scatters correspond to the experimental values deduced from the linear frequency responses at $V_{amp} = 3.16$ mV$_P$, like in Fig. 2; the curves correspond to the natural frequency values reconstructed using the ROM with the identified parameters of TABLE I (solid curves) and the parameters of a hypothetical resonator with 5% thinner resonator (dashed curves). (b) Relative change in the natural frequency predicted by the ROM due to the 5% thickness reduction as a function of $V_{DC}$ for the five different temperatures. (c) Natural frequency predicted by the ROM for $V_{DC} = 0$ V of the identified resonator of TABLE I (solid black curve) and the hypothetical 5% thinner resonator (black dashed curve) as a function of temperature (left y-axis). Relative change in the natural frequency predicted by the ROM due to the 5% thickness reduction as a function of temperature for $V_{DC} = 0$ V (right y-axis).

The large tunability of natural frequency is known to be caused by buckling of the drumhead resonator [39] [30] [27]. To validate this conjecture, we compare the experimental results of Fig. 3a with the predictions of the buckling ROM in Fig. 1c. Therefore, we assign the ROM parameters the values listed in TABLE I from the system-identification algorithm explained in the supplemental material section S3. The solid lines in Figure 3a show the natural frequencies estimated from the ROM for different values of $(T, V_{dc})$. By inspecting Fig. 3a, we conclude that the ROM accurately captures the tunability of the natural frequency as a function of $(T, V_{dc})$ with a relative total error less than 5.5% (cf. supplemental material section S3 for the error definition). We emphasize that these good predictions result from the ROM with parameters independent of the applied temperature and DC voltage (cf. TABLE I). Therefore, the proposed ROM is universal to the applied (temperature, voltage) conditions for the investigated resonator.

For instance, the parameters of the ROM (cf. TABLE I) depend on the geometry/material of the resonator and the residual stresses from fabrication. These effects lead to larger compressive strains at cooler temperatures as shown by the coefficients of $\delta_T$ and $\delta_B$ in TABLE I. In essence, $\delta_T$ and $\delta_B$ monotonically increase (from 0.40 and 0.046 to 0.72 and 0.73, respectively) with the decrease in temperature from 370 K to 295 K (cf. supplemental material Fig. S14a-b). Hence, the ROM indicates that the strong compression in the zero-Volts-state enables the large frequency 2.4× tunability with $V_{DC}$ exhibited in Fig. 3a. Starting from the equilibrium induced by the compressions in the zero-Volts-state, the applied decreases the effective stiffness of the resonator, thus the natural frequency, up to a local minimum (e.g., the decrease in frequency when $V_{DC}$ goes from 4 V to ~17.4 V at 295 K in Fig. 3a). This point of minimum frequency is referred to as the *point of critical buckling* where the structure presents the lowest stiffness due to weak stability [42]. Past the critical buckling value, the applied $V_{DC}$ starts to stretch the resonator, which stabilizes and stiffens the resonator (e.g., the increase in frequency for $V_{DC} > 17.4\ V$ at 295 K in Fig. 3a).

TABLE I. Identified parameters of the ROM in Fig. 1c to fit the experimental results in Fig. 3a for all temperatures and DC voltages with $\kappa_T = 1$, $\bar{d}_T = 1$, and $\bar{d}_E = 10$. Refer to (1), (4), (6), (10), and (14) for the definition of the ROM nondimensional parameters. Refer to the supplemental material section S3 for details about the identification algorithm.

| $\frac{\omega_B}{2\pi}$ [MHz] | $\delta_T$ | | $\delta_B$ | | | $\gamma(\bar{u})$ | |
|---|---|---|---|---|---|---|---|
| | $\theta_0$ | $\theta_1$ [K$^{-1}$] | $\beta_0$ | $\beta_1$ [K$^{-1}$] | $\beta_2$ [K$^{-2}$] | $\gamma_0$ [V$^{-2}$] | $\gamma_1$ [V$^{-2}$] |
| 9.40 | 1.9 | -4.07E-3 | 7.65 | -3.47E-2 | 3.81E-5 | 2.42E-1 | -4.76E-2 |

Remarkably, the smaller temperatures in Fig. 3a not only soften the resonator but also increase its sensitivity to DC voltage. The amplified sensitivity manifests in the abrupt frequency transition around critical buckling at 295 K (i.e., discontinuous slopes). Such abrupt frequency transitions are typically observed in structures with symmetric cross-sections, which are perfectly compliant (i.e., zero natural frequency) at the critical buckling value [42] [34] [43]. Nonetheless, the resonators in this work do not possess symmetric cross-sections (cf. Fig. 1b) and still display the non-smooth frequency-detuning with buckling (cf. Fig. 3a and supplemental material Fig. S2), which enhances the tunability. Therefore, the experiments and the model prove that non-smooth critical buckling can be achieved even in asymmetric structures, resulting in large tunability of the natural frequency.

To highlight the high sensitivity of the dynamics to the resonator's geometry as function of the critical buckling, we use the ROM to examine the effect of a 5% reduction in the thickness of the resonator. By referring to the thickness of the identified resonator in TABLE I as $h_{ID}$, we express a 5% reduced thickness as $h = 0.95\ h_{ID}$. We assume that the thickness $h$ affects only the mass $m$, the stiffnesses $k_B$ and $k_T$ in the ROM of Fig. 1c according to the relationships found in the literature [41] [30] [44] (cf. supplemental material section S4.c for a detailed explanation). These relationships affect the normalized parameters of the ROM listed in TABLE I as follows,

$$\frac{\omega_B}{\omega_{B,ID}} = \frac{h}{h_{ID}} \tag{18a}$$

$$\frac{\kappa_T}{\kappa_{T,ID}} = \left(\frac{h}{h_{ID}}\right)^{-2} \tag{18b}$$

where the "*ID*" subscript denotes the parameter of the identified resonator of thickness $h_{ID}$, and the parameters $\omega_B$ and $\kappa_T$ characterize the resonator with varied thickness $h$.

In Fig. 3a, we overlay the estimated frequencies for the case $h = 0.95\ h_{ID}$, using the scaling of eq.(18). By comparing the solid line (ID) to the dashed line (0.95 ID), we observe that the effect of the thickness variation on the natural frequency is highly dependent on the temperature and voltage of the resonator. This dependence is better visualized in Fig. 3b where we plot the deviation in natural frequency ΔFreq relative to the natural frequency of the identified resonators $Freq_{ID}$. Fig. 3b demonstrates that the magnitude of frequency-deviations diverges close to the critical buckling of each temperature and is larger for the temperatures where the strongest frequency-tunability can be realized. For example, we see a maximum deviation in frequency of -8.5% at 370 K vs. +28.4% at 295 K for the same 5% thickness variation. Hence, a small deviation in parameters of the resonator (like thickness) can strongly affect the vibrations when operating close to the critical buckling. This effect was experimentally and numerically observed in [35] where buckling transition amplifies very small deviations (≤5%) between the unit-resonators in nano-phononic waveguides to the extent of breaking the periodicity of the waveguides and eliminating the acoustic transmission. Note that the scaling (18) ignores the effect of thickness on the residual strain parameters ($\delta_T$ and $\delta_B$) and their temperature-related coefficients (i.e., $\theta_0, \theta_1, \beta_0, \beta_1$, and $\beta_2$). Due to this assumption, the critical buckling occurs around the same $V_{DC}$ for each temperature in the ROM resonators with both thicknesses $h_{ID}$ and $0.95h_{ID}$ as depicted in Fig. 3a. Though, if thickness affects the temperature-related coefficients then the thickness-variations might induce shifts in the value $V_{DC}$ of critical buckling in Fig. 3a, which results in larger frequency-deviations in Fig. 3b.

From the experimental data in Fig. 3a (and the supplemental material Fig. S2), we conclude that achieving the ultra-tunability (e.g., at 320K and 295K) requires a buckling state that is determined by a combination of temperature and voltage effects. For instance, varying the temperature without applying a DC voltage (i.e., $V_{DC} = 0$ V) does not induce a sharp frequency transition as can be predicted by the low $V_{DC}$ data for each temperature value in Fig. 3a. To clarify this behavior, we use the ROM to estimate the frequency of the resonators with thicknesses $h_{ID}$ and $0.95h_{ID}$ as a function of temperature in the absence of DC voltage as shown in Fig. 3c. Indeed, the frequency of the resonators in Fig. 3c tunes smoothly with temperature for $V_{DC} = 0$ V exhibiting a maximum increase of 1.53× and 1.60× for the resonators with $h_{ID}$ and $0.95h_{ID}$, respectively, between 295 K and 370 K. This smooth frequency-tunability realizes a maximum frequency-deviation of -8.2% between the two resonators for the temperature range in Fig. 3c.

### 4. Switching nonlinearity with buckling

Fig. 4a-c show the nonlinear frequency response of the resonator at 340 K, 320 K, and 295 K, respectively. We note that the voltage and temperature not only vary the resonator's natural frequency but also switch the type of nonlinearity in the response. In Fig. 4a at 370 K, for larger oscillation amplitudes, the nonlinear response exhibits stiffening for $V_{DC} = 5, 10$, and 11 V in Fig. 4a(i)-(iii), respectively, but a softening-to-stiffening behavior for $V_{DC} = 15$ V in Fig. 4a(iv). We also observe a switch in nonlinear behavior with $V_{DC}$ at 320 K and 295 K in Fig. 4b-c, where the frequency response shows softening in Figs. 4b(i) and 4c(i) for small $V_{DC}$, stiffening in Fig. 4b(ii)-(iii) and 4c(ii)-(iii) for $V_{DC}$ around the critical buckling transition (cf. Fig. 3a), and finally softening in Fig. 4b(iv) and 4c(iv) for strong $V_{DC}$.

By comparing this switching of nonlinear response in Fig. 4 with the frequency-detuning in Fig. 3a at 370 K, 320 K, and 295 K, we conclude that the nonlinearity is stiffening close to the critical buckling point where the resonator exhibits its softest dynamics due to the applied $V_{DC}$ at the respective temperature. We also observe that the nonlinearity is of softening type far from the critical buckling conditions. This switching trend is clear at 295 K and 320 K where we observe sharp critical buckling points in Fig. 3a, and for 370 K where the natural frequency increases rapidly for $V_{DC} > 11$ V. To summarize, the thermoelastic and electrostatic loads induce *stiffening nonlinearity when the resonator is soft* (i.e., close to the critical buckling point), and *softening nonlinearity when the resonator is stiff* (i.e., away from critical buckling). Moreover, the electrostatic-mediated thermoelastic loads result in a softening-to-stiffening nonlinearity when the resonator is not very far away from critical buckling as seen at 370 K for $V_{DC} = 15$ V in Fig. 4a(iv). We observe this same nonlinearity switching around critical buckling for different resonators at different temperatures and DC voltages as illustrated in the supplemental material Fig. S2-S13 and later in Fig. 5 at 295 K.

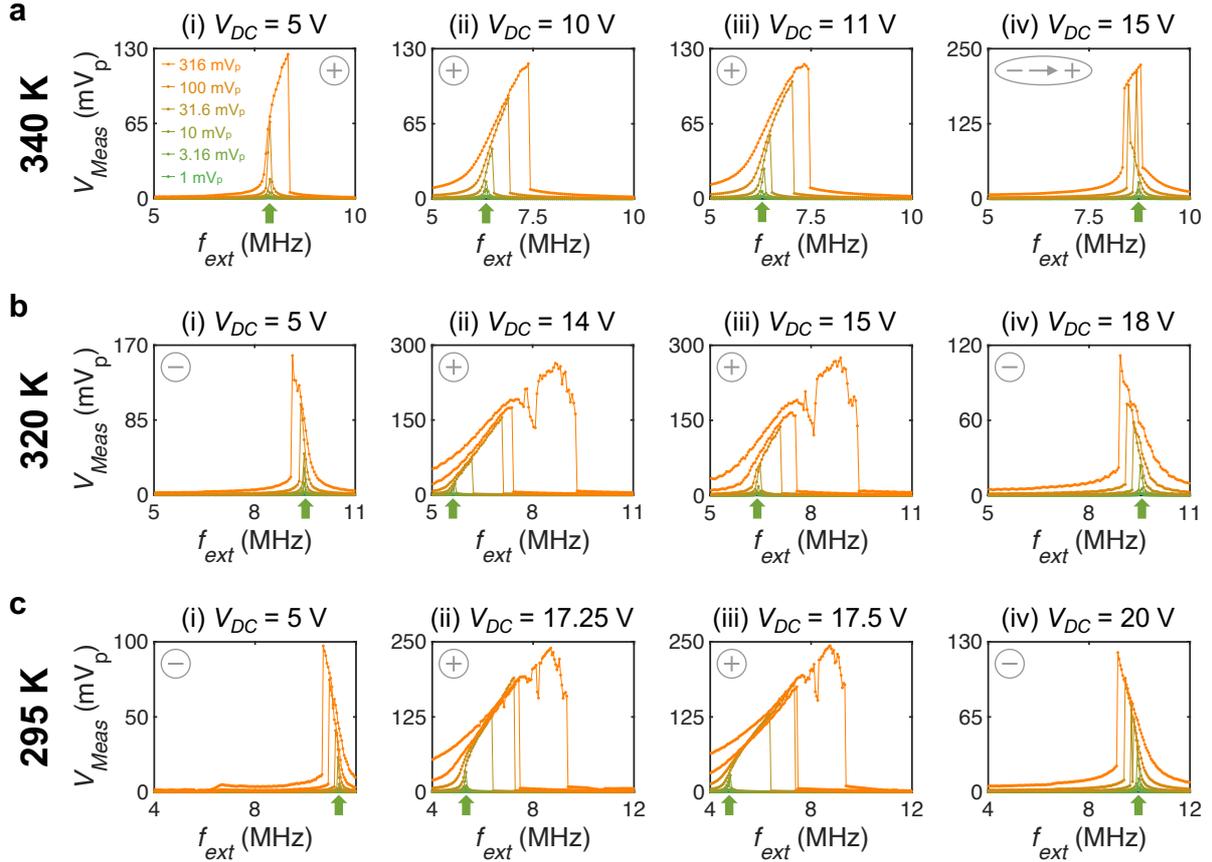

Fig. 4. Electrostatic and thermoelastic tunability of the resonator's nonlinear dynamic response. Experimentally measured frequency response of the resonator at a temperature of (a) 340 K, (b) 320 K, and (c) 295 K for increasing values of $V_{DC}$ (from (i) to (iv)) and $V_{RF}$ (from 1 mV$_p$ to 316 mV$_p$ amplitudes – cf. legend in 2a-i). The voltage $V_{Meas}$ measured by the optical detection apparatus is proportional to the steady-state displacement of the resonator. The plots only show the response in a forward (increasing) sweep of the frequency $f_{ext}$ of the applied $V_{ac}(t) = V_{amp} \cos(2\pi f_{ext} t)$. The voltage unit V$_p$ denotes that the RF voltages are reported using their peak (amplitude) values. The vertical upward arrows point to the linear-resonance frequency at $V_{amp} = 3.16$ mV$_p$, representing the resonator's natural frequency plotted in Fig. 3a. The encircled label in the corner of each subfigure indicates the type of nonlinearity observed at the respective $(T, V_{dc})$ where we refer by "$-$" a softening nonlinearity, "$+$" a stiffening nonlinearity, and "$- \to +$" a softening-to-stiffening nonlinearity.

To explain the physics governing the switching in nonlinearity, we study in Fig. 5 the nonlinear response at 295 K using the ROM of Fig. 1c. We apply (18) to estimate the backbone-curve of the nonlinear frequency response with the parameters of the identified resonator listed in TABLE I. These parameters compute every coefficient in (18) except for the bending nonlinear term $\kappa_3^B$ and the proportionality $\alpha$ relating $\bar{w}_{amp} = \alpha V_{Meas}$ at each temperature $T$ and DC voltage $V_{dc}$. Thus, we identify $\kappa_3^B$ and $\alpha$ for each $V_{dc}$ at 295 K by fitting (18) with $\bar{w}_{amp} = \alpha V_{Meas}$ to the experimental backbone-curve points depicted in Fig. 5a-e. We select these backbone-curve points by choosing the datapoints with maximum amplitude in the continuous responses (i.e., linear responses) and the datapoints at jumps (sudden transitions) in the discontinuous responses (i.e., nonlinear responses).

We plot in Fig. 5f the identified values of $\kappa_3^B$, which reconstructs according to (18) the nonlinear backbone-curves depicted as solid-black lines in Fig. 5a-e. We see that the model captures, (i) the softening nonlinearity away from critical-buckling for $V_{DC} = 5$ V in Fig. 5a, (ii) the softening-to-stiffening nonlinearity around critical-buckling for $V_{DC} = 10$ V, 16 V, and 20 V in Figs. 5b, 5c, and 5e, respectively, and lastly (iii) the stiffening nonlinearity very-close to critical-buckling for $V_{DC} = 17.5$ V in Fig. 5d (cf. Fig. 3a to locate critical-buckling as a function of $V_{DC}$ at 295 K). It is interesting that the model (18) captures this switching in nonlinearity not only qualitatively but also quantitatively by estimating very well the nonlinear frequency-detuning as depicted in Figs. 5a-e (except for $V_{DC} = 10$ V in Fig. 5b where we see some deviation between the experiments and the modeled backbone-curve). Therefore, the ROM in Fig. 1c with the introduced bending nonlinearity $\kappa_3^B$ can accurately model and predict the nonlinear response of the resonator under thermo-electric buckling.

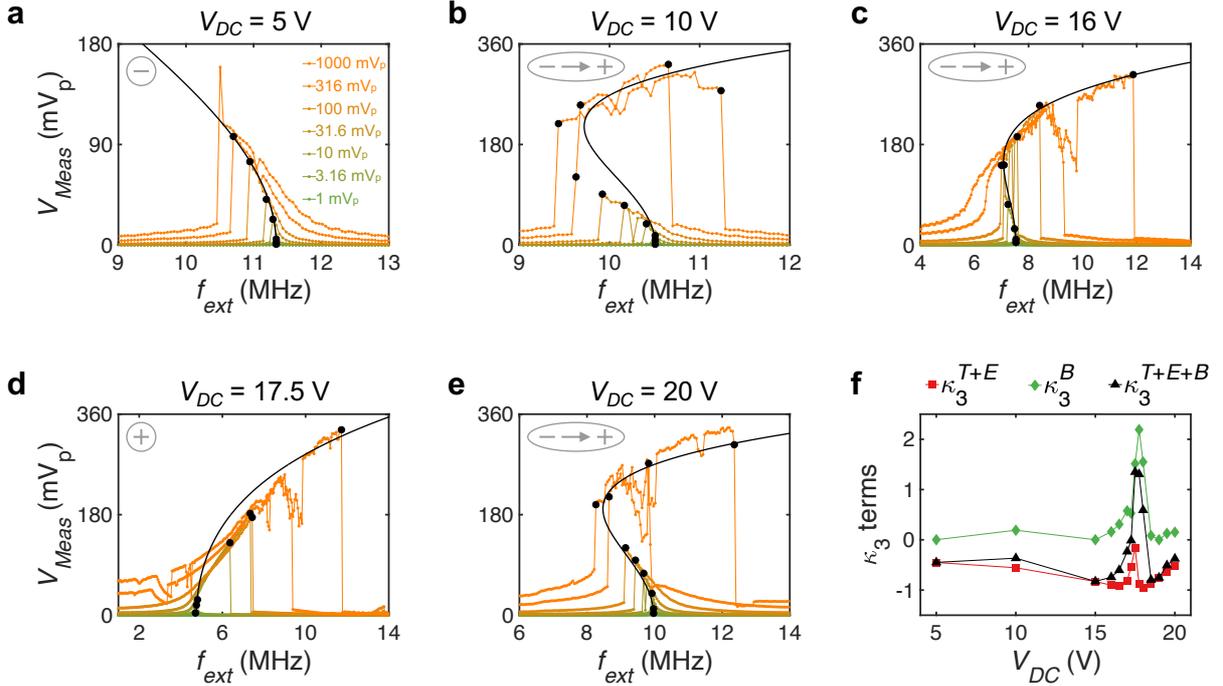

Fig. 5. Electrostatic tunability of the nonlinear response. Experimental frequency response of the resonator at 295 K for $V_{dc}$ of (a) 5 V, (b) 10 V, (c) 16 V, (d) 17.5 V, and (e) 20 V and increasing values of $V_{ac}$ (from 1 mV$_p$ to 1000 mV$_p$ amplitudes – cf. legend in 4a). Black round scatters correspond to the backbone-curve experimental datapoints used to identify $\kappa_3^B$ and $\alpha \stackrel{\text{def}}{=} \frac{\overline{w}_{amp}}{V_{Meas}}$ to reconstruct based on (18) the nonlinear backbone-curves depicted by the solid black lines. The same assumptions and protocols stated in Fig. 2 apply to Fig. 4a-e. The encircled labels in the corner of Fig. 4a-e indicate the type of nonlinearity where we refer by "−" a softening nonlinearity, "+" a stiffening nonlinearity, and "− → +" a softening-to-stiffening nonlinearity. (f) Cubic nonlinear stiffness normalized terms $\kappa_3$ identified in the ROM by the resonator stretching and the electrostatic force $\kappa_3^{T+E}$ (red squares), by the resonator bending $\kappa_3^B$ (green diamonds), and by all the effects $\kappa_3^{T+E+B} = \kappa_3^{T+E} + \kappa_3^B$ (black triangles) as a function of the applied $V_{dc}$.

To explain the physical mechanism of the switching in nonlisnearity, we plot in Fig. 5f the effective cubic nonlinear coefficient $\kappa_3^{T+E}$ induced by stretching and electrostatic effects in addition to $\kappa_3^B$ induced by the bending geometric effect. We investigate the effective cubic nonlinear coefficients (i.e., quadratic, and cubic coefficients) since they govern the nonlinear response for small to moderate amplitudes as dictated by (18). We observe in Fig. 5f that for all the measured values of $V_{DC}$ at 295 K, $\kappa_3^{T+E}$ is negative whereas $\kappa_3^B$ is positive. Hence, the *stretching and electrostatic effects induce softening nonlinearity*, but the *bending geometric effect induces stiffening nonlinearity*.

Moreover, Fig. 5f demonstrates that the stretching and electrostatic effects dominate the total effective cubic nonlinearity $\kappa_3^{T+E+B}$ away from critical buckling (for $V_{DC} \leq 17.25$ V and $V_{DC} \geq 18.5$ V) while the bending geometric effect dominates $\kappa_3^{T+E+B}$ close to critical buckling (for $V_{DC} \in [17.5$ V, $18$ V]). Therefore, the dominance of bending geometric effects at critical buckling drives the switching in nonlinearity from softening to stiffening as the resonator becomes closer to critical buckling in the experiments of Figs. 4, 5a-e, and the supplemental material, cf. Figs. S3-S13. From Fig. 5f, we can also conclude that the stiffening nonlinearity around the critical buckling is (~2 times) stronger than the softening nonlinearity away from the critical buckling. This increase in nonlinearity around critical buckling explains the strong nonlinear frequency-detuning demonstrated by the stiffening response in Figs. 4b(ii)-(iii), 4c(ii)-(iii), and 5d.

## 5. Discussion of the ROM

This ROM serves in modeling the investigated resonators due to its geometry agnosticism. In other words, to apply the ROM, we don't need to know the geometry of the different constituents in the resonators. This property is highly needed for the investigated resonators because they are formed of multiple layers with nonuniform cross-sections [35]. Thanks to this ROM, this work is the first, up to the authors' knowledge, to study buckling in irregular MEMS/NEMS structures. For instance, existing studies treat buckled microresonators by continuum modeling that applies to beams [26] [27] [28] [30] [32] [33] [34] and plates of perfectly rectangular [40] [45] and circular [29] [41] shapes.

However, the continuum modeling allows computing the stresses in the material [27] [30] [34] [40] [41] which our developed ROM can't estimate. Nonetheless, we anticipate that this ROM will find important utility in assisting current analytical [41] [46], numerical [47] [48] [39], and empirical [40] [49] methods that predict the behavior of MEMS structures. In particular, this ROM reduces the degrees of freedom while capturing the essential governing dynamic mechanisms, thus accelerating the design and prototyping processes [47].

## 6. Conclusions

In this work, we experimentally study the electrostatically-mediated thermoelastic tunability of micro-drumhead resonators. We show that these drumhead resonators with high compressive strains – introduced by cooling – present large tunability of their natural frequency with DC voltage. We also find that DC voltage switches the nonlinear response of the resonator by regulating its state of buckling. For instance, the resonators show stiffening nonlinearity, with stronger nonlinear frequency-detuning, close to the critical buckling, while the nonlinearity is of softening type as the resonator moves away from the critical buckling point. We formulated and identified the parameters of a reduced order model (ROM) that captures the experimental linear and nonlinear dynamics for different temperatures and DC voltages. The developed reduced order model is based on lumped mass-springs representations, which offer a valuable tool in designing and studying buckled metamaterials and phononic waveguides.

Indeed, buckling can be exploited to enable new and unusual capabilities in tuning and designing resonators in MEMS. Moreover, the buckling effect can be regulated via electric voltage, which is easily integrated in devices and systems. In this study, optimal buckling performance required also thermal control, which cannot be deployed easily in standalone systems. In the future, additional work should be conducted to design the resonators such that the electric control of buckling leads to ultra-fine tunability at the desired temperature of operation, such as room temperature. Such work requires unraveling the relationship between the structure's mechanics, the strains (e.g., $\delta_T$ and $\delta_B$), temperature, and fabrication-residual stresses. Hence, additional experimental, numerical, and analytical work should be conducted to benefit from buckling for vibrations in NEMS.

## 7. Acknowledgement

This work was supported in part by NSF Emerging Frontiers Research Initiative (EFRI) Grant 1741565. P.F. and A.M.v.d.Z acknowledge support NSF-CAREER Award number CMMI-184673.

# ULTRA-TUNING OF NONLINEAR DRUMHEAD MEMS RESONATORS BY ELECTRO THERMOELASTIC BUCKLING – *SUPPLEMENTAL MATERIAL*

Ali Kanj[1], Paolo F. Ferrari[1], Arend M. van der Zande[1,2],
Alexander F. Vakakis[1], Sameh Tawfick[1,3]

[1]*Department of Mechanical Science and Engineering, University of Illinois at Urbana-Champaign, Illinois 61801, United States*

[2]*Materials Research Laboratory, University of Illinois at Urbana-Champaign, Illinois 61801, United States*

[3]*The Beckman Institute of Advanced Science and Technology, University of Illinois at Urbana-Champaign, Illinois 61801, United States*

This document presents the supplemental material to the article "Ultra Tunable Nonlinear MEMS Resonators Enabled by Electro Thermoelastic Buckling."

## Table of Contents

# S1. Experimental Methods

## a. Resonators Fabrication

We used 4" n++doped silicon wafers from WaferPro™ having (from top to bottom) 100 nm thick ultra-low stress silicon nitride pre-grown by low-pressure chemical vapor deposition (LPCVD) on top of 120 nm silicon dioxide pre-grown by wet oxidation on the base n++ doped silicon wafer. The doped silicon base substrate serves as the ground for electrostatic actuation. The silicon dioxide layer acts as an insulating and sacrificial layer to create the capacitance gap. The silicon nitride forms the drumhead membrane resonator.

To fabricate the actuating electrodes at the end of the waveguides, we pattern the electrodes with a positive resist of AZ5214E using contact photolithography. Then, we sputter 5 nm of chromium and 45 nm of gold. Finally, we define the golden terminals by lift-off in AZ400T of the AZ5214E mask topped with Cr-Au.

Likewise, to pattern the resonators, we use contact photolithography with AZ5214E resist to define the etch holes at the resonator's center. Also, these etch holes allow gas beneath the membranes to escape, which protects the membranes from bursting when measuring under a vacuum. We create these etch holes by Reactive Ion Etching (RIE) in an oxygen and $CF_4$ plasma through the topmost silicon nitride layer.

Afterward, we dice the wafer into 5 mm × 4 mm dies using an automated dicing saw. To suspend the freestanding membranes, we etch the sacrificial silicon dioxide layer with a 10% buffered oxide etchant (BOE) solution, which results in an undercut of the silicon nitride membrane. To drain the etching solution without collapsing the membranes under surface tension forces, we apply supercritical drying resulting in suspended resonators with gold electrodes for actuation, as depicted in Fig. 1a of the main article.

## b. Measurement Details

We attach the fabricated dies to chip carriers via wire ball bonding and place them in a Janus ST-500 optical cryostat under a high vacuum with a less than 5 μTorr pressure. As explained in Fig. 1b of the main article, we wire the doped silicon substrate to the electric ground and the gold terminals to a voltage $V(t) = V_{DC} - V_{RF}(t)$. This in-situ electric configuration allows for electrostatic actuation of the suspended silicon nitride of the resonator. Applying the static voltage $V_{DC}$ and MHz frequency oscillating voltage $V_{RF}$ to the gold electrodes leads to an AC electromechanical z-force $F_{RF} \approx 2(dC/dz)V_{DC}V_{RF}$ where $C$ is the effective capacitance of the membrane, as shown in equation (12) in the main article.

We detect the resulting mechanical vibration with dynamic optical reflectance using Fabry-Pérot interferometry. We point a diffraction-limited 520 nm laser spot to a position close to the resonators' center on the silicon nitride. We avoid pointing the laser close to the edges of the etching holes or the gold terminals. Finally, we measure the dynamic optical reflectance while sweeping the input drive frequency using a network analyzer, resulting in the experimental frequency response plots in Figs. 2 and 4 of the main article.

## S2.   Measurements on Additional Devices

This supplemental document provides additional measurements in Fig. S2-S13, which validate and show the robustness of the experiments presented in the main article. For instance, we observe similar electrostatic-mediated thermoelastic ultra-fine tunability on 3 different resonators located on the same die. We label these resonators as Res. 1, Res. 2, and Res. 3, where Res. 1 is the resonator presented in the main article.

Fig. S2 summarizes the tunability of each resonator's natural (linear) frequency as a function of temperature and DC voltage. Fig. S2 proves that the 3 resonators respond similarly where the maximum tuning with DC voltage depends on the temperature of the resonators. These 3 resonators possess a similar critical-buckling temperature (295 K) because they share similar residual stresses at their location on the same die.

We gathered these natural frequency values for the 3 resonators in Fig. S2 from the frequency response measurements depicted in Fig. S3-S13. Fig. S3-S13 show the tunability of the natural frequency of the resonators and the switching in nonlinear response. By inspecting Fig. S3-S13, we conclude trends in the nonlinearity of the 3 resonators consistent with the conclusions described in the main article: stiffening nonlinearity close to critical buckling and softening nonlinearity away from critical buckling.

## S3.   System Identification Algorithm

In this section, we explain the system identification algorithm used to find the parameters of the model in the main article. In the applied identification, the first step consists of finding the parameters using the experimental detuning of the linear frequency of Fig. 3a (same as Fig. S2a in this document). This step results in all the parameters of the nondimensional force $\bar{F}_{ext}$ in equation (10) of the main article. We use these parameters in the second step to finding the nonlinear bending term $\kappa_3^B$ (cf. equation (18) in the main article) by fitting the experimental backbone curve of the nonlinear frequency responses.

### a. Linear Response

To model the experimental detuning of linear frequency (cf. Fig. 3a and S2), the reduced-order model (illustrated in Fig. 1c in the main article) requires the following set of parameters:

$$P_{all} = \left\{\omega_B, \delta_B(T), \kappa_T, \delta_T(T), \frac{d_T}{d_B}, \gamma_0, \gamma_1, \frac{d_E}{d_B}\right\}. \tag{S1}$$

To reduce the number of parameters to identify, we fix the following parameters:

$$\begin{cases} \kappa_T \stackrel{\text{def}}{=} \dfrac{k_T}{k_B} = 1 \\ \dfrac{d_T}{d_B} = 1 \\ \dfrac{d_E}{d_B} = 10 \end{cases}, \tag{S2}$$

which reduces the identification parameters to:

$$P = \{\omega_B, \delta_B(T), \delta_T(T), \gamma_0, \gamma_1\}. \tag{S3}$$

The decision to assign values to the parameters in (S2) reduces the number of parameters, facilitating the identification. We select the values in (S2) based on a physical estimate that the stretching and bending effects possess similar order of magnitude in the membrane (values for $\kappa_T$ and $\frac{d_T}{d_B}$), and that the electric gap is 10 times larger than $d_B$ the size of the effective bending spring (value of $\frac{d_E}{d_B}$). We note that the assigned values in (S2) do not define the stretching and electric force values. The exact magnitudes of stretching and force values depend on the parameters $P$ in (S3). Therefore, (S2) is only a decision to reduce the number of parameters in $P_{all}$ to the effective parameters governing the resonator dynamics. We validate (S2) by observing the ability of the model to capture the experimental measurements, which results in a good agreement (with ~5.11% relative error) as shown in Fig. 3a in the main article.

To investigate the modeling of $\delta_B$ and $\delta_T$ as a function of temperature $T$, we expand the parameter set in (S3) to:

$$P = \left\{\omega_B, \begin{array}{l} \delta_B(370\text{ K})\ \delta_T(370\text{ K}) \\ \delta_B(360\text{ K})\ \delta_T(360\text{ K}) \\ \delta_B(340\text{ K}), \delta_T(340\text{ K}), \gamma_0, \gamma_1 \\ \delta_B(320\text{ K})\ \delta_T(320\text{ K}) \\ \delta_B(295\text{ K})\ \delta_T(295\text{ K}) \end{array}\right\}. \tag{S4}$$

By fitting the experimental frequency $f_{Exp}$ in Fig. 3a for all the applied temperatures and DC voltage, we identify the values for the parameters in (S4). To perform the fitting, we first compute the stable static equilibrium position $\bar{u}_{EQM}$ for each $(T, V_{DC})$ according to expression (11) in the main article using the function "fsolve" in MATLAB®. Then, we find the linear/natural frequency at the respective $\bar{u}_{EQM}$ by evaluating:

$$f_{Model} = \frac{\omega_B}{2\pi} \frac{d\bar{F}_{ext}}{d\bar{u}}\bigg|_{\bar{u}=\bar{u}_{EQM}}, \qquad (S5)$$

where $f_{Model}$ is the estimated natural frequency based on the ROM at $(T, V_{DC})$.

In this fitting (i.e., system identification), we search for the set of parameters in (S3) that minimizes:

$$Error \stackrel{\text{def}}{=} \frac{1}{\min f_{Exp}} \sqrt{\frac{1}{N} \sum (f_{Exp} - f_{Model})^2}, \qquad (S6)$$

where $N$ is the number of $f_{Exp}$ values and $\min f_{Exp}$ is the minimum frequency over the entire measured $(T, V_{DC})$ domain in Fig. 3a. We search for the minimum error (S6) using the function "patternsearch" in MATLB®, which occurs for the values of $\omega_B$, $\gamma_0$, and $\gamma_1$ listed in Table I of the main article and the values of $\delta_T$ and $\delta_B$ presented in Fig. S14a-b, respectively. With these optimal parameters, the model predicts the experimental results with an error (S6) of ~4.37%, as depicted in Fig. S14c.

Hence, substitution of these parameters into the model provides accurate fit as shown in Fig. 3a in the main article, where the error (S6) is ~5.11%. This can be attributed to the fact that $\delta_T$ and $\delta_B$ in Fig. S14c are optimal at every temperature given by Fig. S14a-b. Whereas in Fig. 3a, $\delta_T$ and $\delta_B$ follow temperature-dependent models (cf. equations (6) in the main article) that do not lead to the exact optimal values in the case of Fig. S14c. For instance, we identified the temperature-dependent models in equations (6) in the main article by fitting $\delta_T$ and $\delta_B$ to polynomial fits as indicated in Fig. S14a-b, respectively. These fittings result in the coefficients $\theta_0$, $\theta_1$, $\beta_0$, $\beta_1$, and $\beta_2$ listed in Table I of the main article.

### b. Nonlinear Response

In addition to the parameters listed in Table I, we need two new parameters, namely $\kappa_3^B$ and $\alpha \stackrel{\text{def}}{=} \frac{W_{amp}}{V_{Meas}}$ to compute the nonlinear frequency detuning $\sigma$ according to equation (18) in the main article. For this purpose, we identify $\kappa_3^B$ and $\alpha$ for every $V_{DC}$ at 295 K by fitting equation (18) to the backbone curve experimental points, as shown in Fig. 4a-e of the main article.

For this fitting, we also apply the function "patternsearch" in MATLAB® resonator at 295 K and each $V_{DC}$ with the error function:

$$Error(V_{DC}) \stackrel{\text{def}}{=} \frac{1}{f_{Exp\;Lin}(V_{dc})} \sqrt{\frac{1}{N} \sum (\sigma_{Exp} - \sigma_{Model})^2}, \quad (S7)$$

where $f_{Exp\;Lin}(V_{DC})$ is the experimental linear/natural frequency of the (cf. of Fig. 3a), $\sigma_{Exp} = f_{Exp\;Nonlin} - f_{Exp\;Lin}(V_{DC})$ with $f_{Exp\;Nonlin}$ being the frequency of the experimental backbone curve points, and $\sigma_{Model}$ is the nonlinear detuning modeled by equation (18) in the main article. These fittings result in the values of $\kappa_3^B$ presented in Fig. 4f for every $V_{DC}$ at 295 K, and model the detuning as exemplified in Fig. 4a-e.

## S4. Model Considerations Based on The Experimental Identifications

In this section, we highlight some of the assumptions that we adopted based on the system identification of the model.

### a. Modeling The Effect of Bending on the Electrostatic Force

In equation (7) in the main article, we express the electric force $F_E$ in terms of a surface $S$ corresponding to the effective overlay between the resonator gold pads and the ground Si substrate (cf. Fig. 1). Referring to Fig. S15a, this surface $S$ can be derived from:

$$F_E(t) = -\frac{\epsilon_0 V^2}{2} \iint_{S_{Gold}} \frac{dxdy}{[d_E + \tilde{u}(x,y,t)]^2}, \quad (S8)$$

where we denote by $F_E$ the total electrostatic force applied at the resonator (oriented positively away from the grounded substrate), $\epsilon_0$ the permittivity of free space, $V$ the voltage applied between the substrate and the resonator, $S_{Gold}$ the surface of the resonator gold pads (cf. Fig. 1), $d_E$ the baseline gap between the substrate and the resonator, and $\tilde{u}(x,y,t)$ the out-of-plane deformation of the resonator's membrane at location $(x,y)$ and instant $t$.

The reduced-order model (ROM) assumes that the resonator deforms similarly to its fundamental out-of-plane mode, which mathematically writes into:

$$\tilde{u}(x,y,t) \cong u(t) \cdot [1 - U(x,y)] \quad (S9)$$

where $u(t)$ is the displacement of the mass in the ROM as illustrated in Fig. 1c of the main article and $U(x,y)$ is the shape of the fundamental out-of-plane mode. We illustrate this assumption in Fig. S15a where $u(t)$ corresponds to the maximum displacement of the membrane. With the ROM assumption in (S9), (S8) becomes:

$$F_E(t) = -\frac{\epsilon_0 V^2}{2[d_E + u(t)]^2} \overbrace{\iint_{S_{Gold}} \left[1 - \frac{u(t)U(x,y)}{d_E + u(t)}\right]^{-2} dxdy}^{S}. \tag{S10}$$

We are interested in the case of $u(t) \ll d_E$ resulting in the linear approximation:

$$\left[1 - \frac{u(t)U(x,y)}{d_E + u(t)}\right]^{-2} \approx \left[1 - \frac{u(t)U(x,y)}{d_E}\right]^{-2} \approx 1 + 2\frac{u(t)U(x,y)}{d_E}. \tag{S11}$$

Applying (S11) in (S10), we get:

$$F_E(t) \approx -\frac{V^2}{[d_E + u(t)]^2} \overbrace{\left[\overbrace{\frac{\epsilon_0}{2} \iint_{S_{Gold}} dxdy}^{c_0} + \overbrace{\frac{\epsilon_0}{d_E} \iint_{S_{Gold}} U(x,y)dxdy \, u(t)}^{c_1}\right]}^{\epsilon_0 S/2} \tag{S12}$$

identical to equation (8) modeling $F_E$ in the main article where $S$ represents an effective capacitance surface.

This model in (S12) considers the linear change in the effective capacitance surface $S$ due to the bending of the membrane. It is important to incorporate (at least) this linear change to avoid modeling errors. For instance, if we ignore the membrane's bending and assume a parallel-plates model as illustrated in Fig. S15b, $U(x,y) = 0$ in equation (S10), leading to:

$$F_E(t) = -\frac{V^2}{[d_E + u(t)]^2} \cdot \overbrace{\frac{\epsilon_0}{2} \iint_{S_{Gold}} dxdy}^{c_0}. \tag{S13}$$

With this parallel-plates model in (S13), the system identification (as explained in section S3a) at every temperature leads to the results presented in Fig. S15b-c.

Fig. S15b shows the optimal values of $\gamma_E \stackrel{\text{def}}{=} \frac{c_0}{2k_B d_B^3}$ at each temperature, which exhibit large variations. To ensure the universal applicability of the model over the entire temperature domain, we used the average of these $\gamma_E$'s to compute the linear frequency of the ROM with the parallel-plates model for every temperature and DC voltage in Fig. S15c. We note that this parallel-plates model does not perform well in modeling the experimental detuning of the natural frequency. Therefore, it is crucial to consider the bending of the membrane when modeling the electrostatic force, like our ROM approach in the main article.

### b. Neglecting Parametric Excitation

Equation (15) in the man article indicates that the applied RF voltage induces external and parametric excitations. However, in the main article, we ignored the parametric excitation because its amplitudes ($\Gamma_n$) are negligible compared to the spring coefficients ($\kappa_n$). To validate this assumption, we compare $|\Gamma_1|$ vs. $\kappa_1$, $|\Gamma_2|$ vs. $\kappa_2$, and $|\Gamma_3|$ vs. $\kappa_3^{T+E}$ in Fig. S16a-c, respectively, for all the experimental DC and RF voltages. The plots in Fig. S16 confirm our assumption to ignore the parametric excitation.

### c. Modeling The Effect of Thickness on The ROM Parameters

In equations (19) in the main article, we present scaling laws that estimate the change of $\omega_B$ and $\kappa_T$ with the resonator thickness. According to [1], the flexural characteristic frequency of circular membrane scales proportionally to the thickness, leading to equation (19a) in the main article.

As for equation (19b), recall that $\kappa_T \stackrel{\text{def}}{=} \frac{k_T}{k_B}$ is a measure of stretching rigidity to flexural rigidity in the membrane. The flexural rigidity of the membrane scales cubically with the thickness [1]; whereas the stretching rigidity scales linearly with the thickness [2]. Therefore, we assumed that $\kappa_T$ scales inversely with the square of the thickness as indicated by equation (19b) in the main article.

## S6. Figures

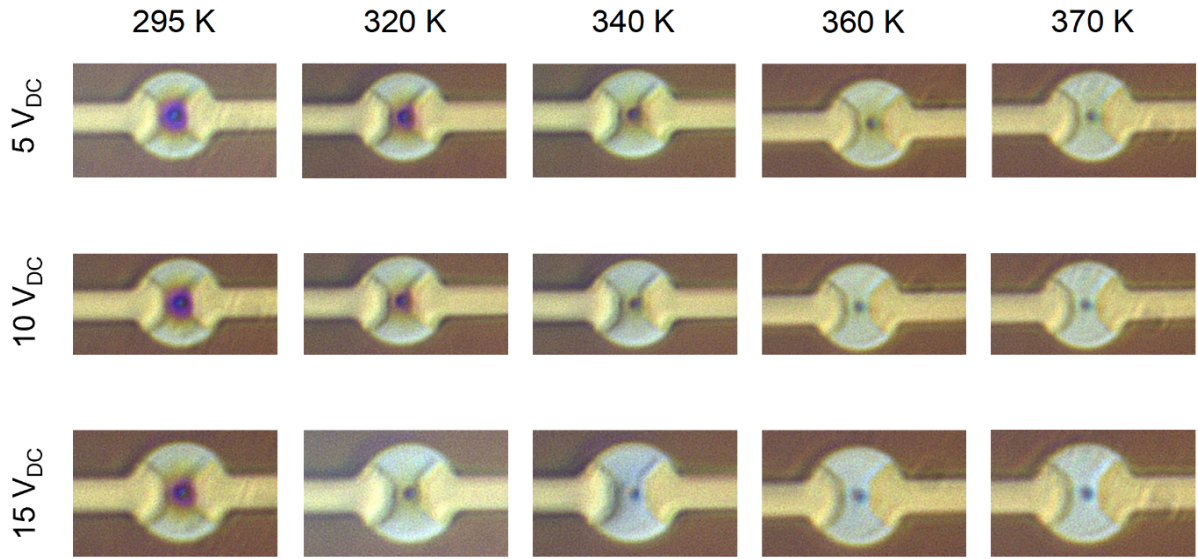

Fig. S1. Optical micrograph of the investigated resonator (Res. 1) at 295, 320, 340, 360, and 370 K under $V_{DC}$ = 5 V (top row), 10 V (middle row), and 15 V (bottom row). The resonator diameter is ~14 μm. The purple color at the center of the resonator indicates that the center is deflected compared to the circumference, which results in the purple color due to the optical interferences. We note that the strong deflections (e.g., 295 and 320 K) correspond to the cases with strong effect of buckling.

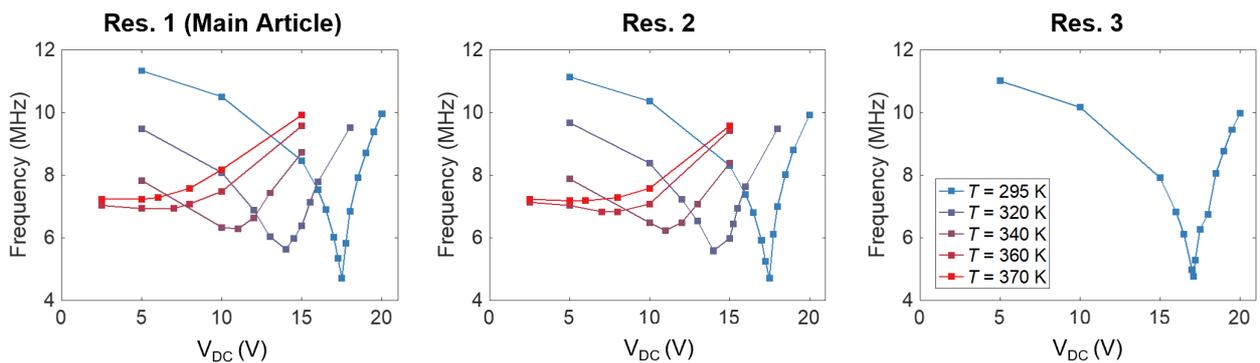

Fig. S2. The natural frequencies of three different resonators as a function of the applied $V_{DC}$ for five different temperatures: Res. 1 (featured in the main article), Res. 2, and Res. 3 (from left to right). The square scatters correspond to the experimental values deduced from the frequency responses with $V_{amp}$ = 3.16 mV$_p$ in Fig. S3-S13.

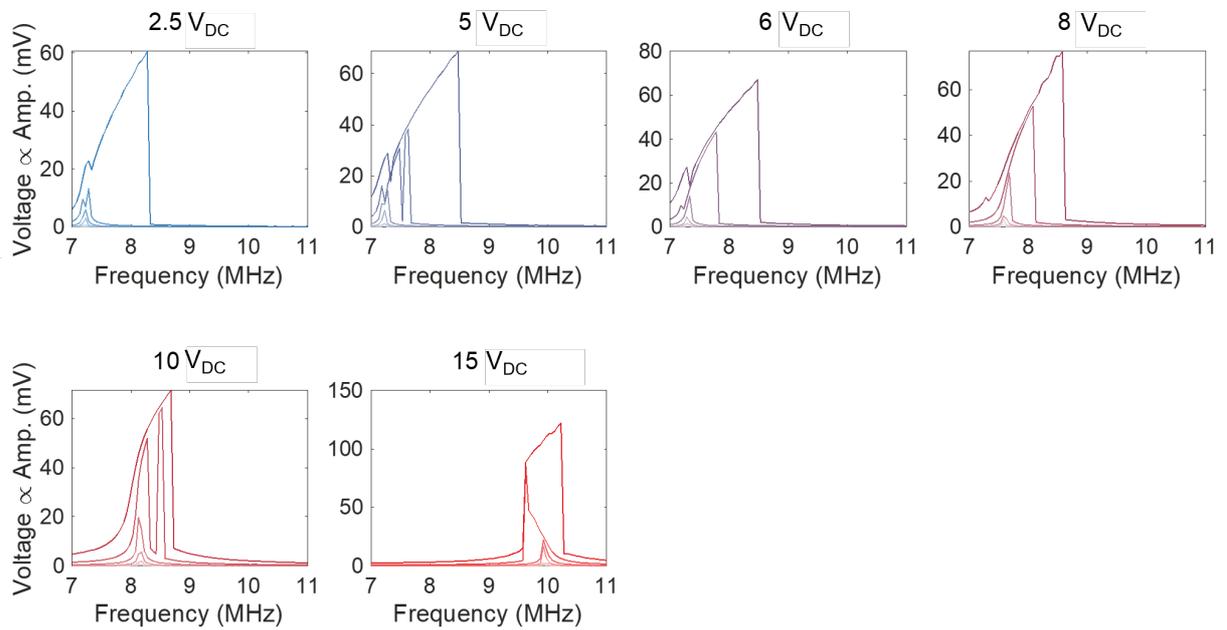

Fig. S3. Frequency response measurements of Res. 1 (featured in the main article) at 370 K.

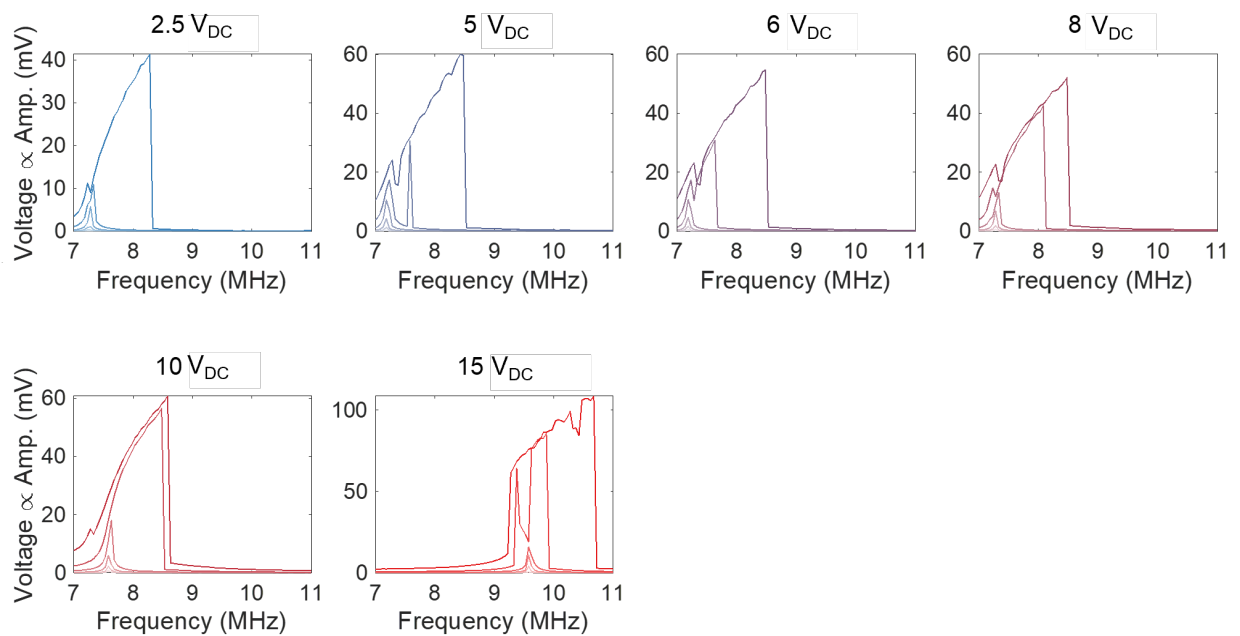

Fig. S4. Frequency response measurements of Res. 2 at 370 K.

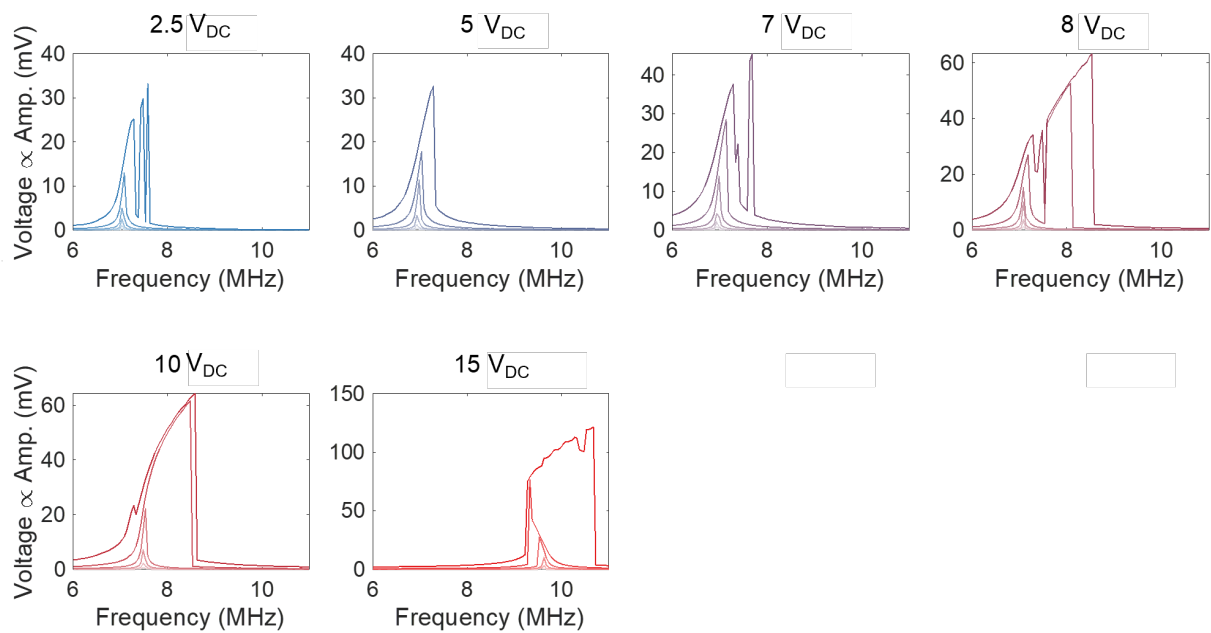

Fig. S5. Frequency response measurements of Res. 1 (featured in the main article) at 360 K.

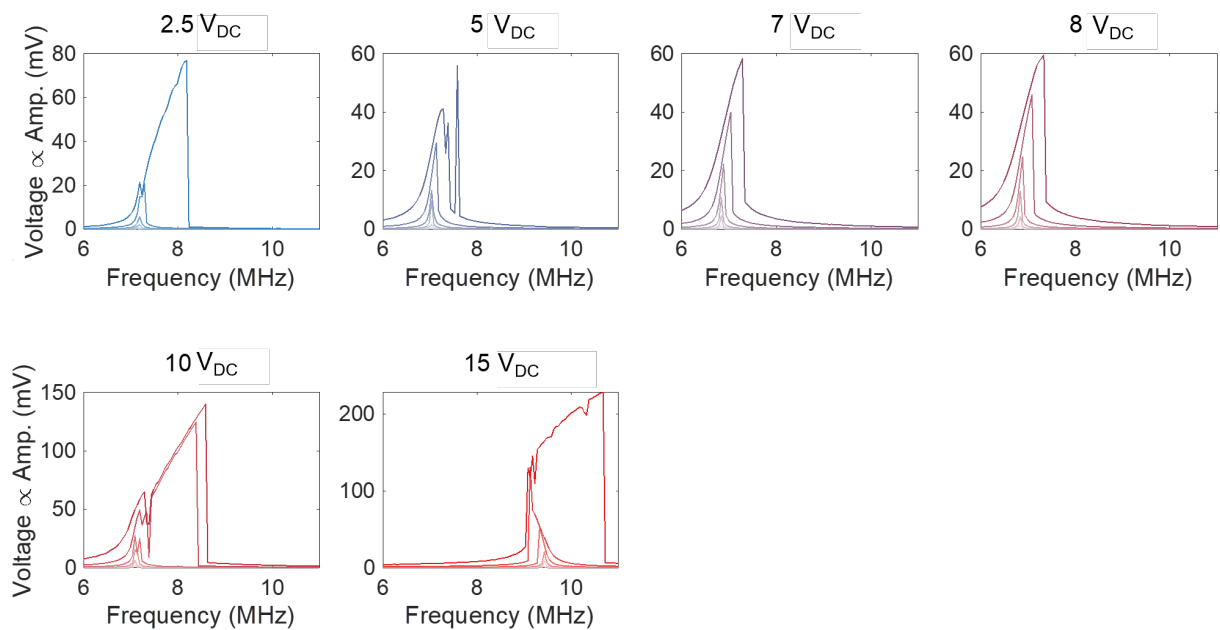

Fig. S6. Frequency response measurements of Res. 2 at 360 K.

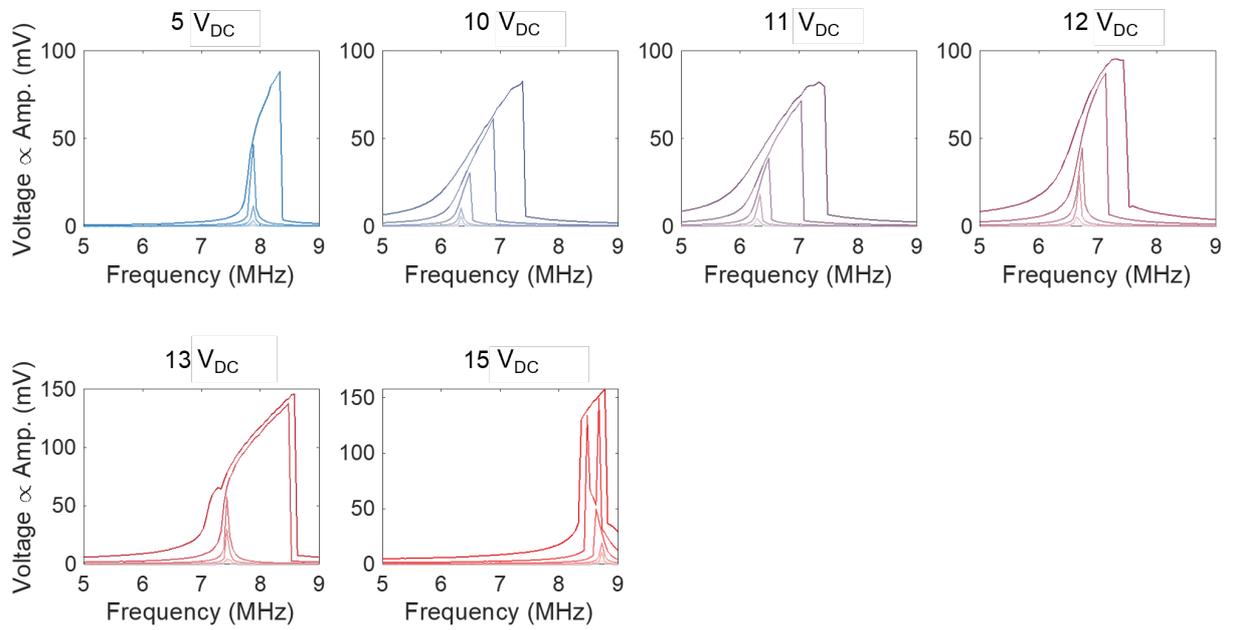

Fig. S7. Frequency response measurements of Res. 1 (featured in the main article) at 340 K.

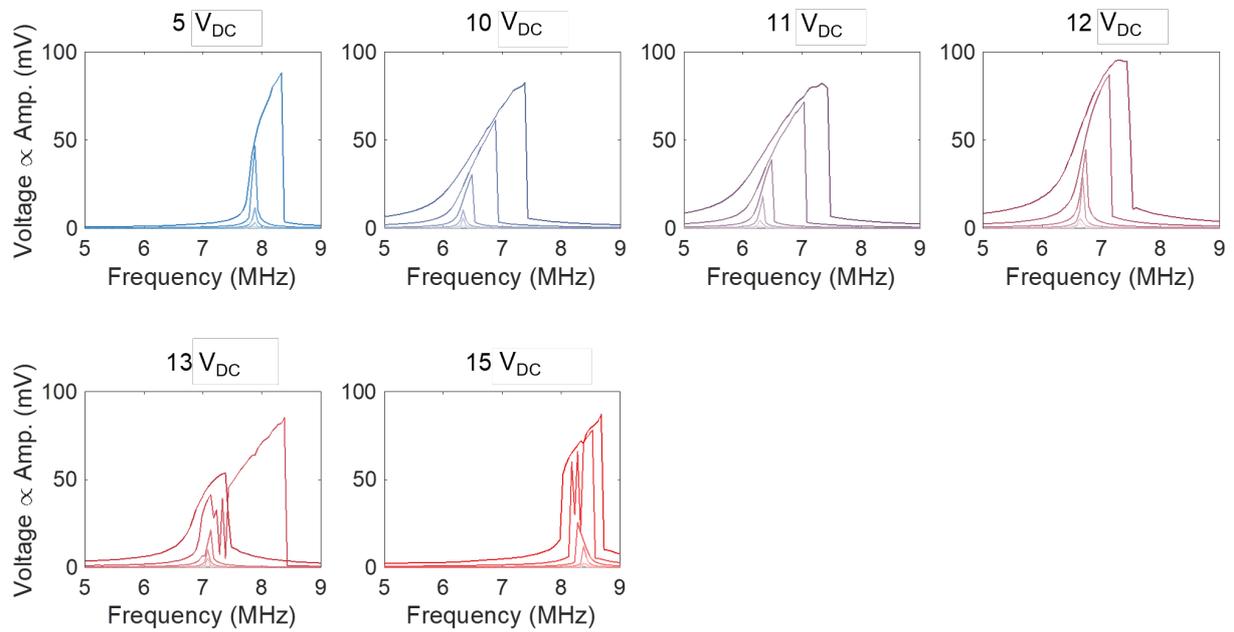

Fig. S8. Frequency response measurements of Res. 2 at 340 K.

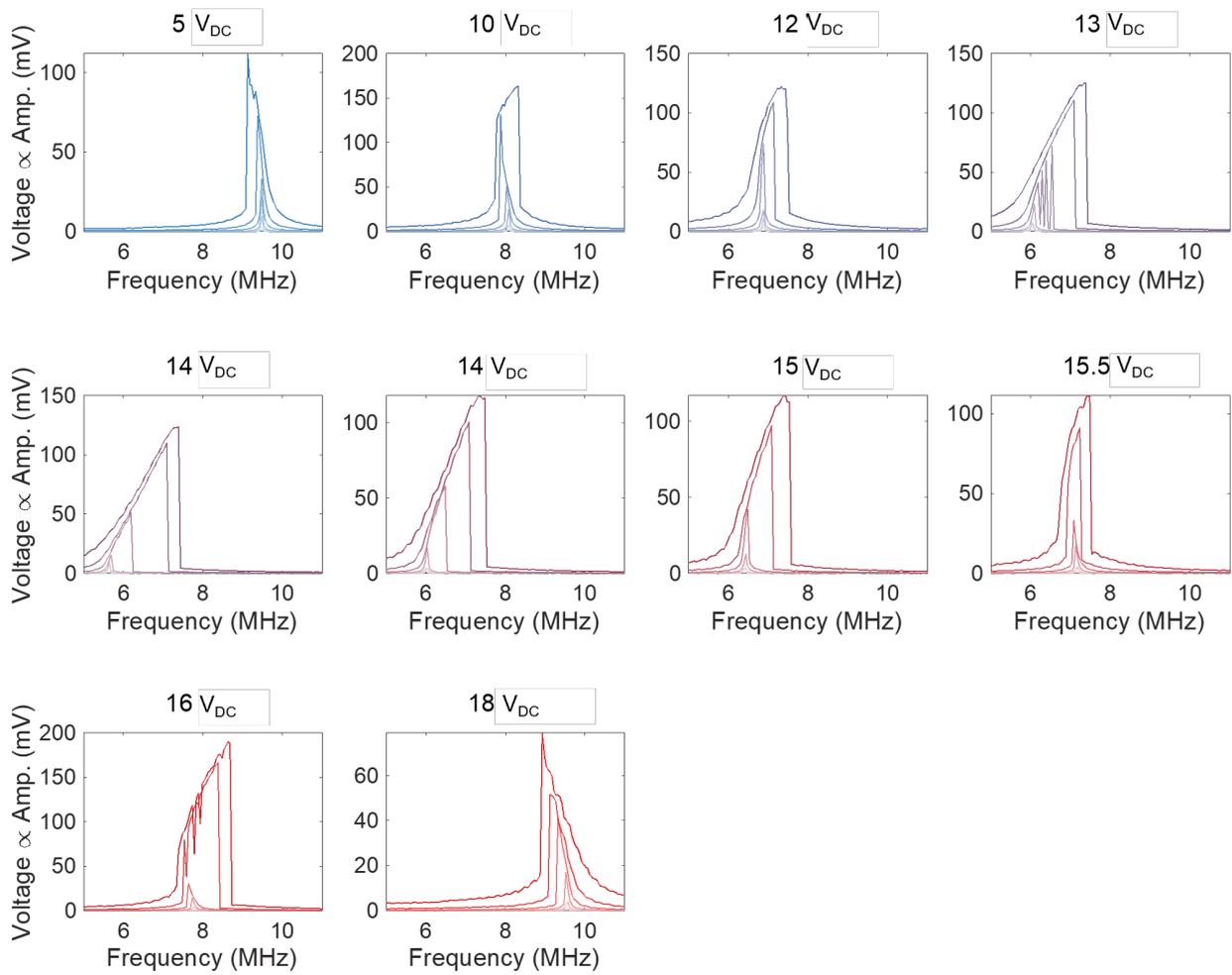

Fig. S9. Frequency response measurements of Res. 1 (featured in the main article) at 320 K.

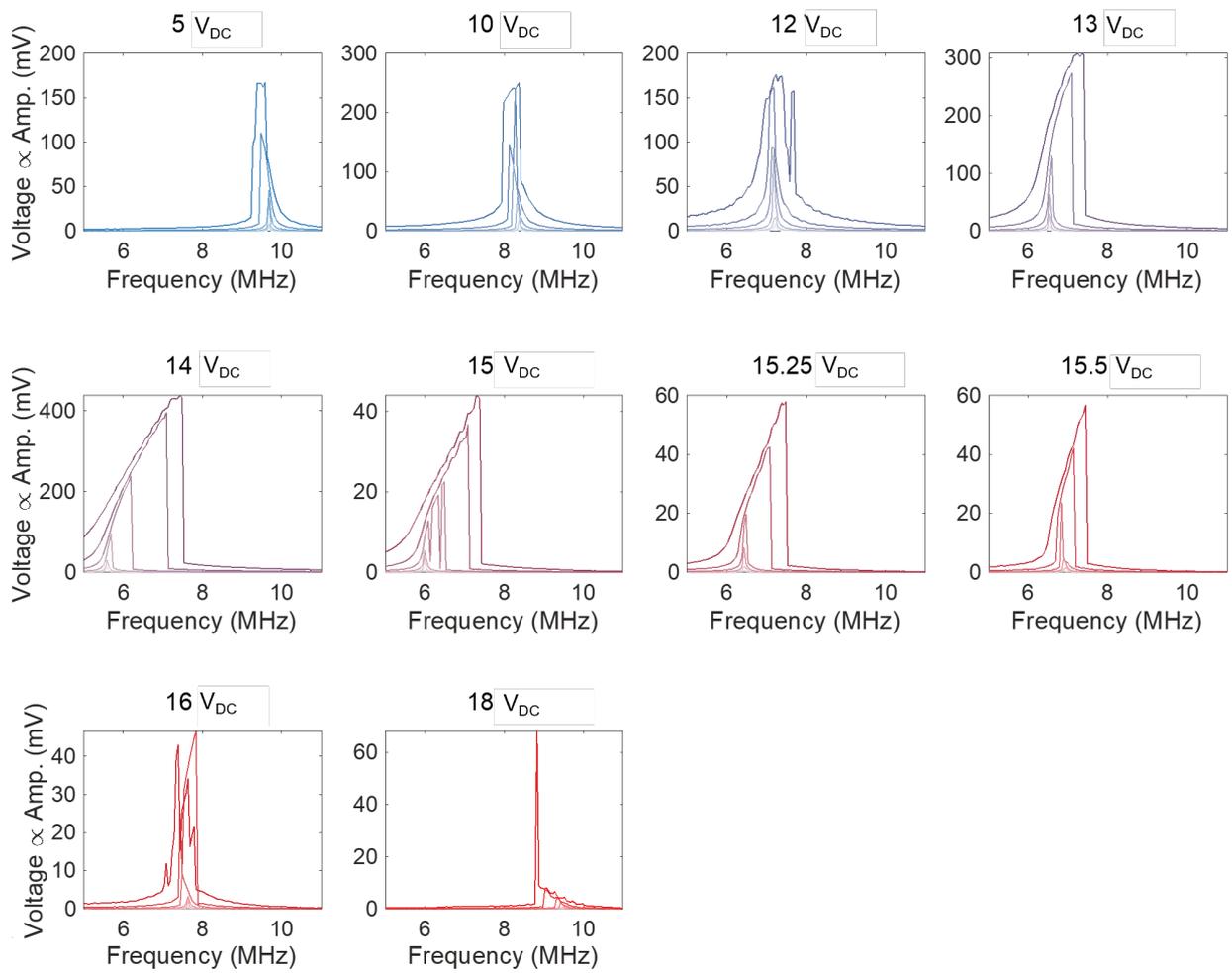

Fig. S10. Frequency response measurements of Res. 2 at 320 K.

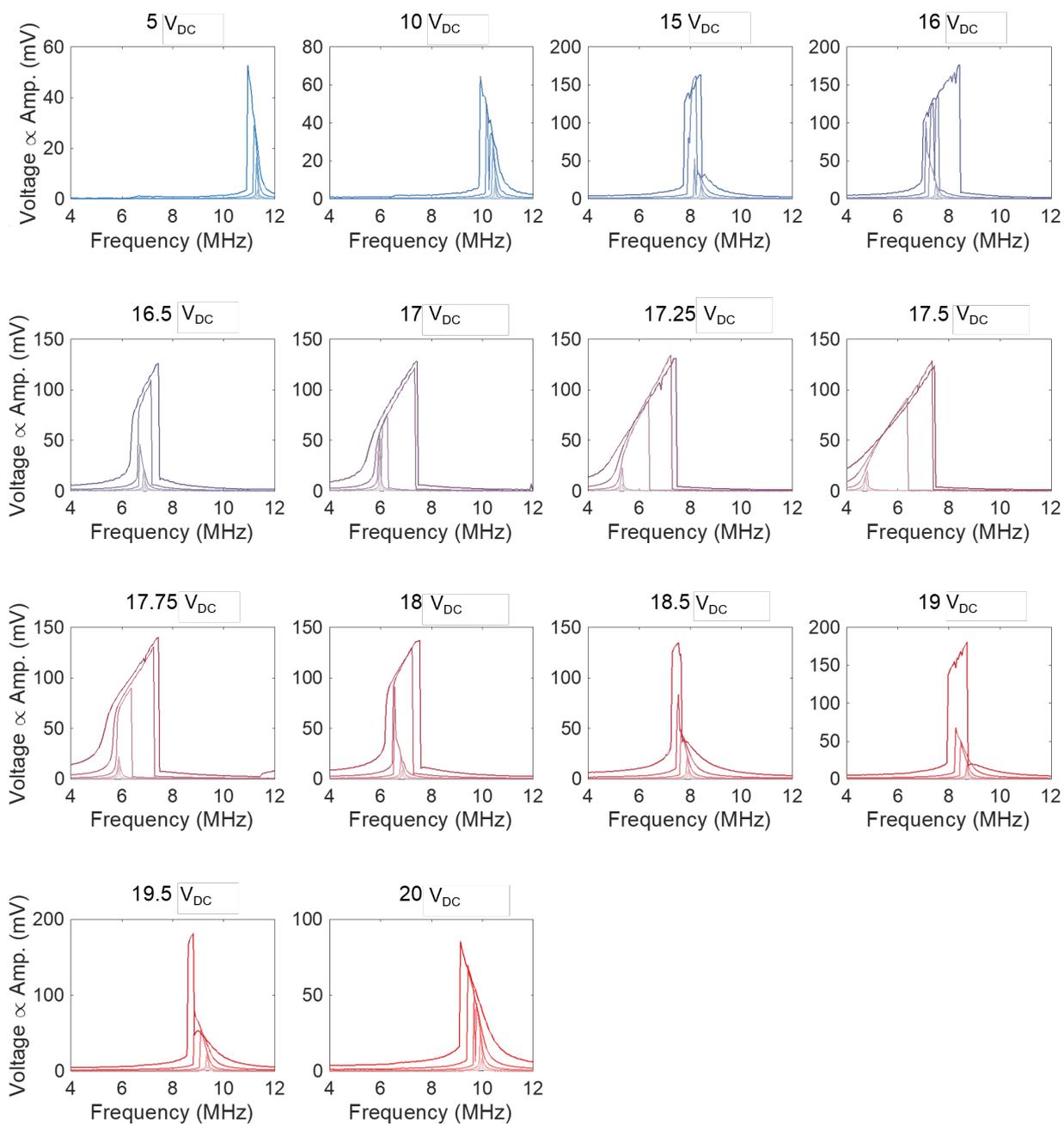

Fig. S11. Frequency response measurements of Res. 1 (featured in the main article) at 295 K.

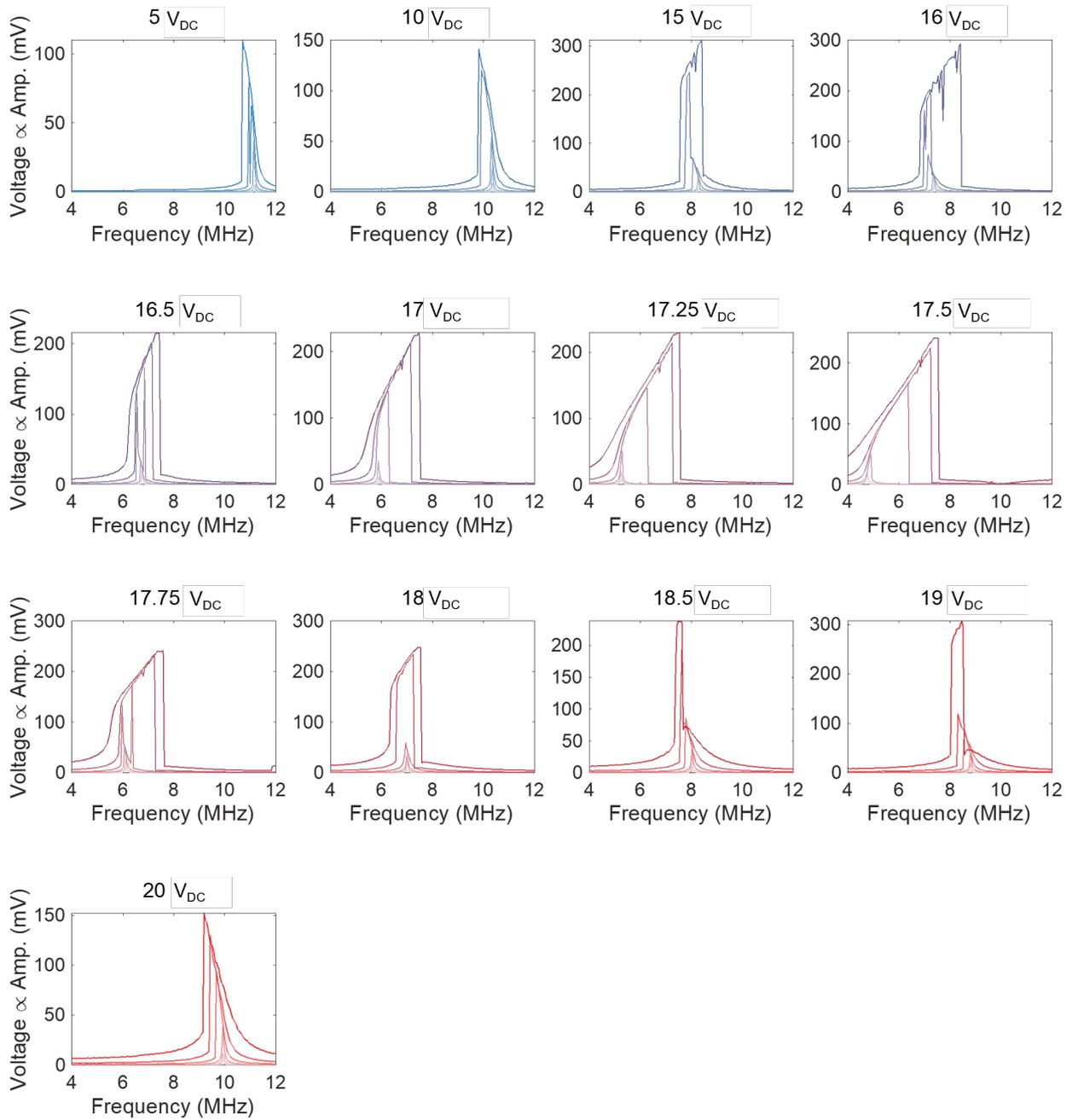

Fig. S12. Frequency response measurements of Res. 2 at 295 K.

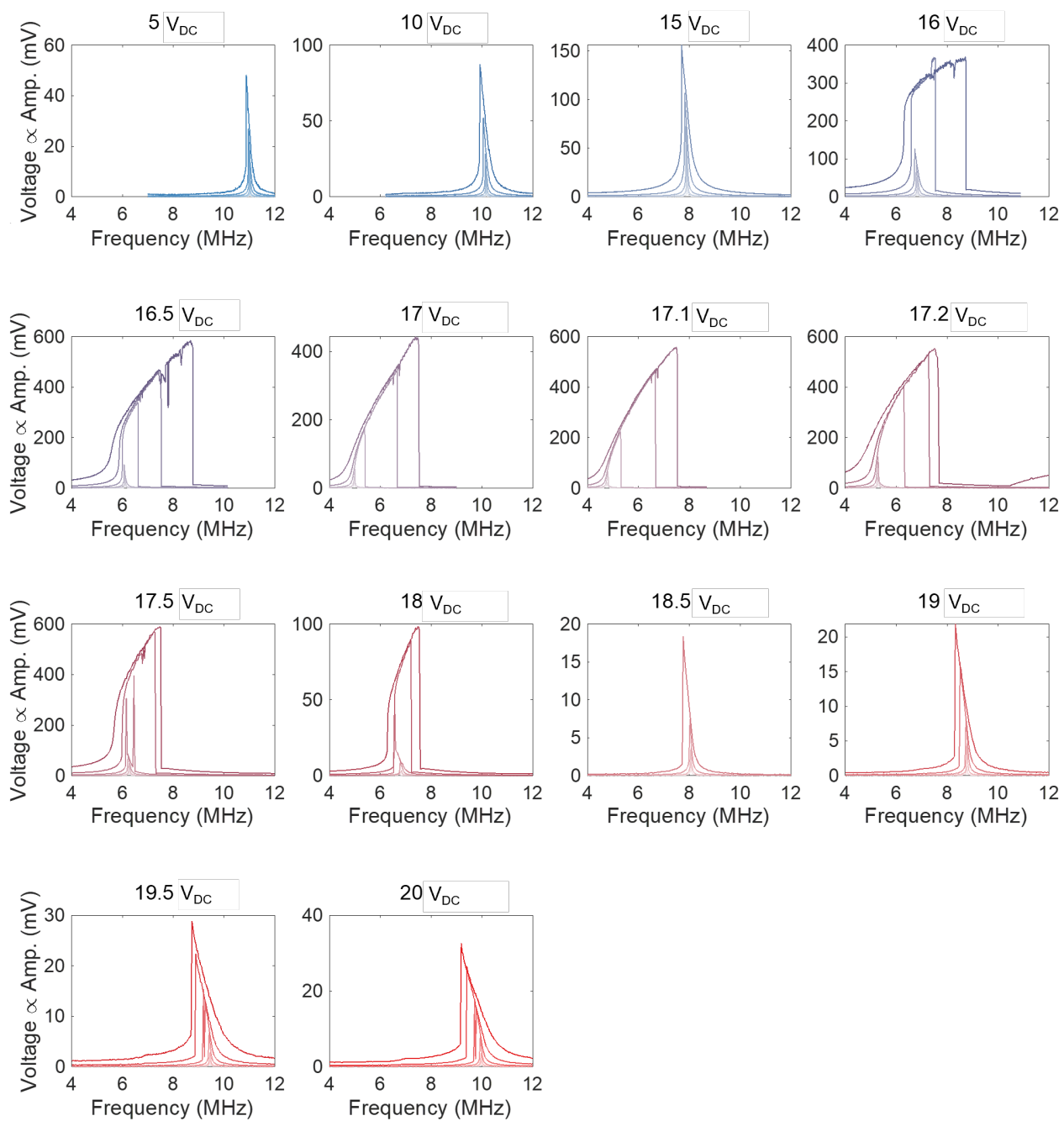

Fig. S13. Frequency response measurements of Res. 3 at 295 K.

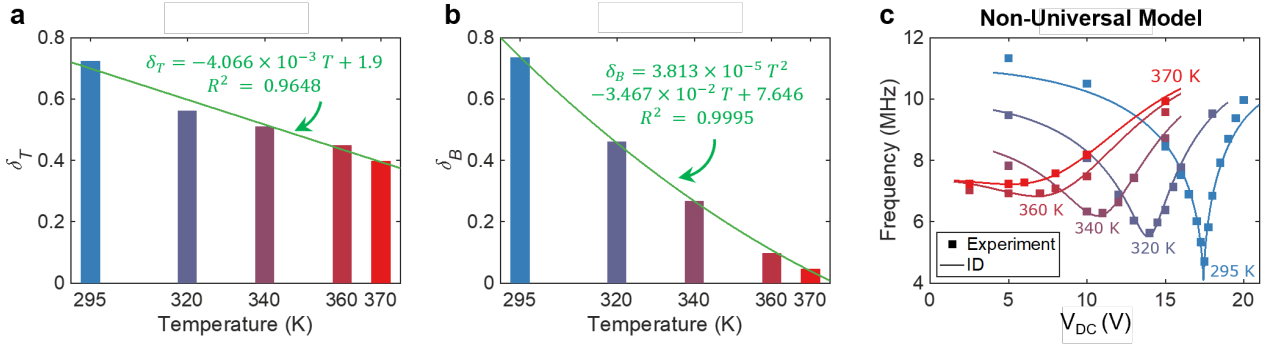

Fig. S14. Identified values of (a) $\delta_T$ and (b) $\delta_B$ as a function of temperature based on the algorithm described in section S3a. The green curves correspond to the polynomial fits and the resulting $R^2$. (c) Reconstruction of the natural frequencies of the resonator as a function of the applied $V_{dc}$ and temperatures of Res. 1 (featured in the main article) using the values of $\delta_T$ and $\delta_B$ identified at each temperature. Hence, this model is not "universal" because $\delta_T$ and $\delta_B$ are defined locally at each temperature. For this reason, this model presents overfitting (~4.37%) compared to the universal model in Fig. 3a (~5.11% error). The square scatters correspond to the experimental values deduced from the frequency responses of Res. 1 with $V_{amp}$ = 3.16 mV$_p$.

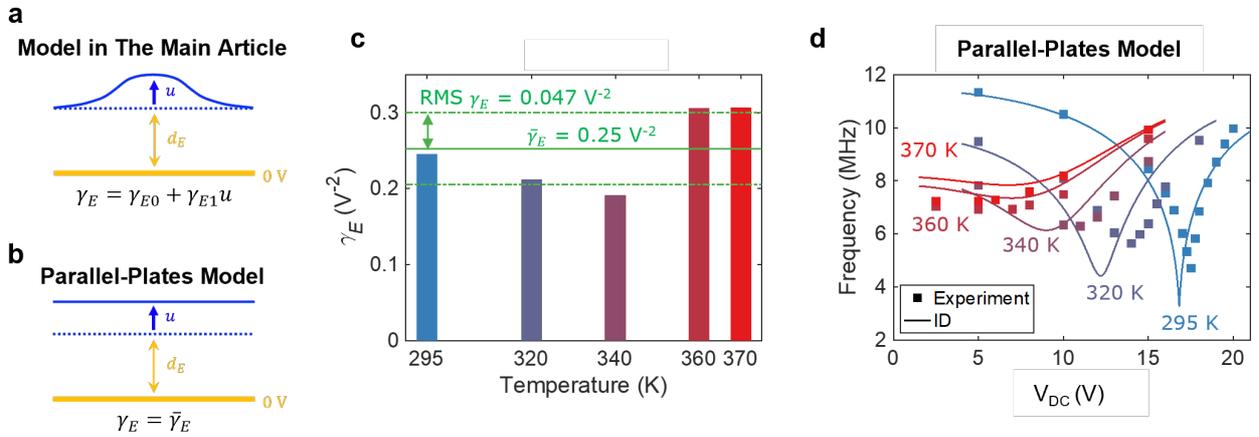

Fig. S15. Schematics illustrating the change in capacitor-gap due to deflection $u$ at the center of the resonator under (a) the bending model adopted in the main article and (b) the parallel-plates model. (c) Identified values of $\gamma_E$ as a function of temperature based on the algorithm described in section S3a. The solid green curve corresponds to the mean value of $\gamma_E$ and the dashed green curves correspond to the boundaries within one RMS value $\approx \pm 0.047$ V$^{-2}$ of the mean $\gamma_E$. (d) Reconstruction of the natural frequencies of the resonator as a function of the applied $V_{DC}$ and temperatures of Res. 1 (featured in the main article) using the parallel-plates model with the mean $\gamma_E$ of Fig. S15c. Note the bad performance of the parallel-plate model in estimating the frequencies compared to the model in the article, as depicted in Fig. 3a.

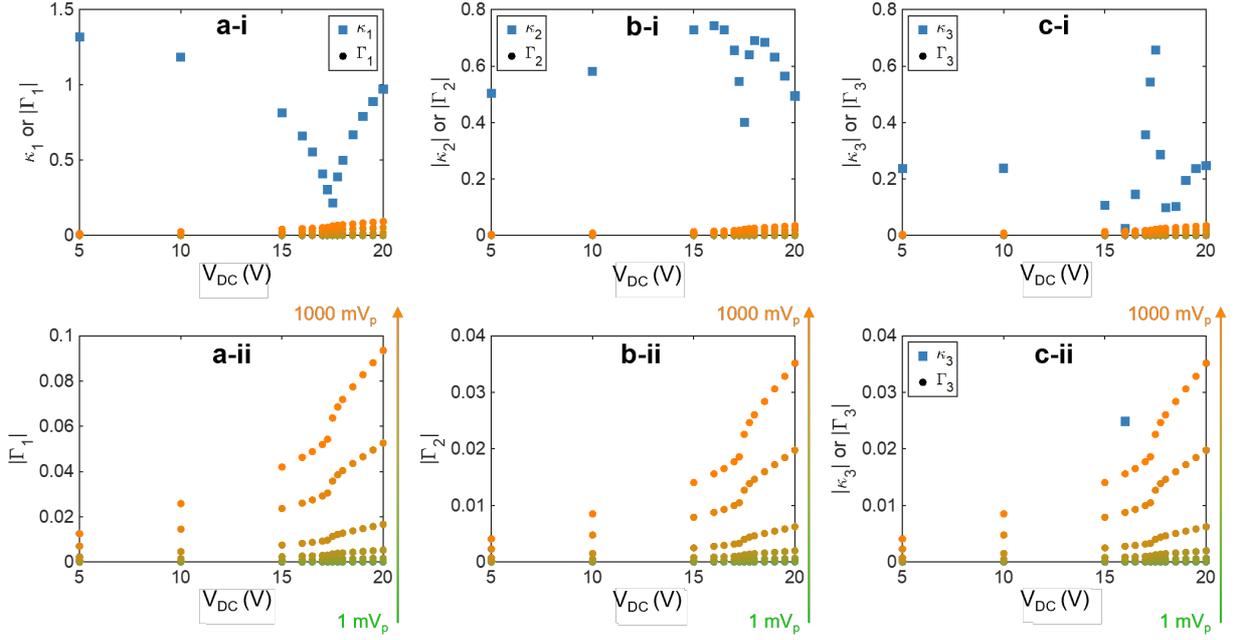

Fig. S16. Comparison at 295 K of Res. 1 (featured in the main article) between the (tension-bending-electric) the stiffness coefficients $\kappa_n \stackrel{\text{def}}{=} \frac{1}{n!} \frac{d^n \bar{F}_{DC}}{d\bar{u}^n}\Big|_{\bar{u}=\bar{u}_{EQM}}$ and the parametric excitation amplitude $\Gamma_n \stackrel{\text{def}}{=} \frac{2}{n!} V_{dc} V_{amp} \frac{d^n \Gamma}{d\bar{u}^n}\Big|_{\bar{u}=\bar{u}_{EQM}}$ for order (a) 1, (b) 2, and (c) 3. The bottom row (i.e., figures ii) shows a y-zoomed view of $\Gamma_n$ which are computed for $V_{amp}$ = 1, 3.16, 10, 31.6, 100, 316, and 1000 mV$_p$. Note that for $n = 1, 2$, and 3, the stiffness coefficients are much higher than the parametric excitation amplitude even though we only consider the stiffness nonlinearity due to tension and electrostatics (i.e., $\kappa_3^B$ is not included in Fig. S16c).

19